\documentclass[11pt]{article}
\pdfoutput=1
\usepackage{jheppub}

\usepackage{graphicx}
\usepackage{dcolumn}
\usepackage{bm}
\usepackage{color}
\usepackage{epstopdf}

\def\rn{\rho}
\def\rha{\rho_5}
\def\rh{r_h}
\def\x{\hat{x}}
\def\br{(r)}
\def\d{\mathrm{d}}
\def\p{\partial}
\def\ve{\varepsilon}

\def\0{}
\def\bo{}

\def\hr{{\hat{r}}}
\def\homega{\hat{\omega}}
\def\hmu{\hat{\mu}}
\def\chiV{\chi_{_V}}
\def\chiA{\chi_{_A}}

\def\Fv{F}
\def\Fa{{\tilde{F}}}
\def\Av{A}
\def\Aa{{\tilde{A}}}

\def\Up{q}
\def\tUp{\tilde{q}}
\def\qV{q_{_V}}
\def\qA{q_{_A}}

\def\tQ{\tilde{Q}}
\def\tT{{\tilde{T}}}
\def\av{a}
\def\aa{{\tilde{a}}}
\def\davi{\delta {\av}_i^{(0)}}
\def\daai{\delta {\aa}_i^{(0)}}
\def\dhti{\delta h_{ti}^{(0)}}

\def\R{\mathcal{R}}
\def\S{\mathcal{S}}
\def\T{\mathcal{T}}

\def\K{\mathcal{K}}
\def\M{\mathcal{M}}
\def\E{\mathcal{E}}

\def\P{\mathcal{P}}
\def\Q{\mathcal{Q}}
\def\Lc{{\gamma}}
\def\o{{O}}

\def\mv{\mu}
\def\ma{\mu_5}

\def\Em{E}

\def\svq{\sigma_{e}}
\def\saq{\sigma_{5e}}
\def\sv{\sigma}
\def\sa{\sigma_{5}}
\def\stq{{\tilde{\sigma}}_{e}}
\def\staq{{\tilde{\sigma}}_{5e}}
\def\st{\tilde{\sigma}}
\def\sQ{\sigma_Q}

\def\sto{\tilde{\sigma}_{(\omega)}}
\def\svo{\sigma_{(\omega)}}
\def\sao{\sigma_{5(\omega)}}
\def\avo{\alpha_{(\omega)}}
\def\aao{\alpha_{5(\omega)}}
\def\bavo{\bar\alpha_{(\omega)}}
\def\bkvo{\bar\kappa_{(\omega)}}

\def\rme{(\epsilon)}
\def\bs{(u)} 
\def\seed{(u)}
\def\corr{(1)}
\def\eq{(0)}

\def\Ew{{W}}
\def\Ev{{W}}
\def\Ea{{\tilde{ {W}}}}
\def\nn{\nonumber}
\def\({\left(}
\def\){\right)}
\def\[{\left[}
\def\]{\right]}

\title{Holographic Charged Fluid with Chiral Electric Separation Effect} 

\author[a]{Yanyan Bu,}
\author[b]{Rong-Gen Cai,}
\author[c]{Qing Yang,}
\author[d]{Yun-Long Zhang}

\affiliation[a]{Department of Physics, Harbin Institute of Technology, Harbin 150001, China}

\affiliation[b]{CAS Key Laboratory of Theoretical Physics, Institute of Theoretical Physics,\\
Chinese Academy of Sciences, Beijing 100190, China and \\ School of Physical Sciences,
University of Chinese Academy of Sciences, Beijing 100049, China}

\affiliation[c]{Department of Astronomy, Beijing Normal University, Beijing 100875, China}

\affiliation[d]{Asia Pacific Center for Theoretical Physics, Pohang 790-784, Korea and\\
Center for Quantum Spacetime, Sogang University, Seoul 121-742, Korea}

\emailAdd{yybu@hit.edu.cn}
\emailAdd{cairg@itp.ac.cn}
\emailAdd{yangqing@bnu.edu.cn}
\emailAdd{yunlong.zhang@apctp.org}

\abstract{
Hydrodynamics with both vector and axial currents is under study within a holographic model, consisting of canonical $U(1)_V\times U(1)_A$ gauge fields in an asymptotically AdS$_5$ black brane. When gravitational back-reaction is taken into account, the chiral electric separation effect (CESE), namely the generation of an axial current as the response to an external electric field, is realized naturally. Via fluid/gravity correspondence, all the first order transport coefficients in the hydrodynamic constitutive relations are evaluated analytically: they are functions of vector chemical potential  $\mu$, axial chemical potential $\mu_5$ and the fluid's temperature $T$. Apart from the proportionality factor $\mu\mu_5$, the CESE conductivity is found to be dependent on the dimensionless quantities $\mu/T$ and $\mu_5/T$ nontrivially. As a complementary study, frequency-dependent transport phenomena are revealed through linear response analysis, demonstrating perfect agreement with the results obtained from fluid/gravity correspondence.
}

\arxivnumber{1803.08389}
\preprint{\href{https://doi.org/10.1007/JHEP09(2018)083}{https://doi.org/10.1007/JHEP09(2018)083}}
\dedicated{September 27, 2018} 

\begin{document}

\maketitle
\allowdisplaybreaks

\section{Introduction} \label{intro-sum}

Fluid dynamics is an effective low energy description of most interacting systems at finite temperature. Within such a hydrodynamic approximation, the entire dynamics of a microscopic theory is reduced to that of macroscopically conserved currents, such as the stress-energy tensor and charge current operators computed in a locally near equilibrium thermal state. An essential ingredient of any fluid dynamics is the constitutive relations which relate the macroscopically conserved currents to the hydrodynamic variables, such as fluid velocity and charge densities, and to external forces like external electromagnetic fields. Derivative expansion in the hydrodynamic variables and external forces accounts for deviations from thermal equilibrium. At each order, the derivative expansion is fixed by thermodynamics and symmetries, up to a finite number of transport coefficients such as viscosity and diffusion coefficients. These transport coefficients are not calculable from hydrodynamics itself but have to be deduced experimentally or computed from the underlying microscopic theory.

For relativistic fluid dynamics, the stress-energy tensor is conveniently parameterized as
\begin{equation} \label{stress-cov}
T_{\mu\nu}=\E u_\mu u_\nu + \P P_{\mu\nu}+ \pi_{\mu\nu},
\end{equation}
where $u_\mu,\E,\P$ are the velocity, energy density and pressure of the fluid, and  $P_{\mu\nu}=u_\mu u_\nu +\eta_{\mu\nu}$ is the projection tensor. Up to the first order in the derivative expansion, the viscous component $\pi_{\mu\nu}$ takes the form,
\begin{equation} \label{pimunu}
\pi_{\mu\nu}=-\eta P_\mu^\alpha P_\nu^\beta\big(\partial_\alpha u_\beta + \partial_\beta u_\alpha -\frac{2}{3} P_{\alpha\beta} \partial_\sigma u^\sigma\big) -\zeta P_{\mu\nu} \partial_\alpha u^\alpha +\mathcal{O}(\partial^2),
\end{equation}
where $\eta$, $\zeta$ are the shear viscosity and bulk viscosity, respectively. Throughout this work we will take the Landau-Lifshitz frame so that $u^\mu\pi_{\mu\nu}=0$.

For the fluid with conserved charges, one also needs to specify constitutive relations for the associated currents. Indeed, the charge transport properties are found to be useful in probing the structure and dynamics of matter. A well-known example is the Ohm's law $J^i \equiv\langle \bar\psi\gamma^i \psi \rangle= \sigma  E^i$, which states the generation of an electric current in response to an external electric field for a normal conducting media. In recent years, exploration of other possible electric current generation, particularly for a system with charged chiral fermions, has attracted much interest. It turns out that the celebrated microscopic chiral anomaly induces fascinating anomalous transport phenomena that break the space parity symmetry. One such example is the chiral magnetic effect (CME) \cite{Kharzeev:2004ey,Kharzeev:2007tn,Kharzeev:2007jp,Fukushima:2008xe}: generation of an electric current directed along an externally applied magnetic field, $\vec{J}={\xi_B} \vec{B}$. The existence of CME relies on chirality imbalance between left- and right-handed chiral fermions, usually parameterized by an axial chemical potential $\mu_5$. For a rotating hydrodynamic flow, chiral anomaly induces a vector current along the fluid vorticity, $\vec{J}=\frac{1}{2}\xi \vec{\nabla}\times \vec{u}$, which is called the chiral vortical effect (CVE) \cite{Banerjee:2008th,Erdmenger:2008rm,Son:2009tf}.

On the other hand, an axial current $J^i_5 \equiv\langle \bar\psi\gamma^i \gamma_5 \psi \rangle$ also exists for a system with charged chiral fermions. In fact, via the chiral anomaly effect, an axial current is generated along an external magnetic field, $\vec{J}_5= {\xi_{5B}} \vec{B}$, which is referred to as chiral separation effect (CSE) \cite{Son:2004tq,Metlitski:2005pr}. Interestingly, the interplay of CME and CSE predicts a gapless wave mode propagating along the magnetic field, called chiral magnetic wave (CMW) \cite{Kharzeev:2010gd}. In heavy-ion collisions, the CMW induces an electric quadrupole moment of the quark-gluon plasma \cite{Burnier:2011bf}, leaving experimentally observable effects \cite{Kharzeev:2010gr,Burnier:2011bf,Bzdak:2012ia,Yee:2013cya}.
It is important to stress that chiral anomaly-induced transport phenomena summarized above are non-dissipative  \cite{Son:2009tf}, that is they do not contribute to entropy production. While observable signatures predicted by the CME have not been conclusively detected in heavy-ion collisions \cite{Adam:2015vje,Khachatryan:2016got, Sirunyan:2017quh,Sirunyan:2017tax}, the CME may explain a large negative magneto-resistance observed in Dirac and Weyl semi-metals \cite{Huang:2015eia,Li:2016,Li:2014bha}. We recommend recent reviews \cite{Kharzeev:2013ffa,Huang:2015oca,Kharzeev:2015znc,Skokov:2016yrj,Landsteiner:2016led,Gorbar:2017lnp} and references therein on the subject of anomalous transport phenomena.

The separation of chiral charge could also be induced by an external electric field, $\vec{J}_5= \saq \vec{E}$, which is called chiral electric separation effect (CESE) \cite{Huang:2013iia,Jiang:2014ura}. However, it is important to emphasize that the CESE does not originate from chiral anomaly but is simply the result of conduction of a chiral many-body environment \cite{Huang:2013iia}. On very general grounds, the CESE exists only when both vector and axial chemical potentials are nonzero. Based on Kubo formula, the CESE conductivity has been computed for thermal quantum electrodynamics (QED) \cite{Huang:2013iia} and quantum chromodynamics (QCD) plasmas \cite{Jiang:2014ura} at high-temperature regime up to leading-log order in gauge couplings. In the strongly-coupled regime, the CESE was studied in \cite{Pu:2014cwa, Pu:2014fva} by using the holographic QCD model of \cite{Sakai:2004cn,Sakai:2005yt}.
Recently, the CESE conductivity is also computed within the framework of kinetic theory \cite{Gorbar:2016qfh,Gorbar:2017vph, Gorbar:2018vuh, Gorbar:2018nmg} under the relaxation time approximation (RTA). If the CESE conductivity is parameterized as $\saq=\mu \mu_5 \tilde{\chi}_A$, then all these studies indicate that $\tilde{\chi}_A$ depends on $\mu,\mu_5$ weakly. In analogy with CMW, the CESE, combined with the Ohm's law, results in a new gapless wave mode propagating along the electric field, which is called density wave \cite{Huang:2013iia} or chiral electric wave (CEW) \cite{Pu:2014fva}.

The transport phenomena reviewed above could be summarized into hydrodynamic constitutive relations for vector and axial currents. In the Landau-Lifshitz frame where $u_\mu J^\mu =-\rho$ and $u_\mu J^\mu_5 =- \rho_5$, we are ended up with \cite{Sadofyev:2010pr, Neiman:2010zi, Kalaydzhyan:2011vx},
\begin{align}
{J}^\mu &={\rn}\, u^\mu +\svq \Big[ \Em^\mu -   {T}  P^{\mu\nu} \partial_\nu \Big(\frac{\mv}{T}\Big) \Big] - \staq {T}   P^{\mu\nu} \partial_\nu \Big(\frac{\ma}{T}\Big)+\(\xi \omega^\mu+{\xi_B} B^\mu\)+\mathcal{O}(\partial^2),   \label{covariantJm} \\
{J}^\mu_5&={\rn}_5\, u^\mu +\saq \Big[\Em^\mu {-}  {T}  P^{\mu\nu} \partial_\nu \Big(\frac{\mv}{T}\Big) \Big] {-} {\stq} {T}  P^{\mu\nu} \partial_\nu \Big(\frac{\ma}{T}\Big)+\(\xi_5\omega^\mu+{\xi_{5B}} B^\mu\) +\mathcal{O}(\partial^2), \label{covariantJm5}
\end{align}
where $\rho,\rho_5$ are vector and axial charge densities. The external electromagnetic fields $E^\mu,B^\mu$ and the fluid's vorticity $\omega^\mu$ are
\begin{equation}
E^\mu\equiv F_{\rme}^{\mu\nu} u_\nu, \qquad      B^\mu\equiv\frac{1}{2} \epsilon^{\mu\nu\alpha\beta} u_\nu F^{\rme}_{\alpha\beta},     \qquad \omega^\mu \equiv \frac{1}{2} \epsilon^{\mu\nu\alpha\beta} u_\nu \partial_\alpha u_\beta.
\end{equation}
In \eqref{covariantJm}\eqref{covariantJm5}, the Wiedemann-Franz law has been used to relate thermal conductivity to electrical conductivity. The $\tilde{\sigma}_{5e}$- and $\tilde{\sigma}_{e}$-terms are relevant to chiral charge diffusions, while $\xi_5$-term is an axial analogue of CVE. Time/space evolution of the system is determined by solving the hydrodynamic equations
\begin{align} \label{hydro eqs}
\partial^\mu {T}_{\mu\nu}={J}^{\alpha} F^{\rme}_{\nu\alpha} ,\qquad \quad
\partial_\mu {J}^\mu=0,\qquad \quad
\partial_\mu {J}^\mu_5=\mathcal{C} E_\mu B^\mu ,
\end{align}
where the axial current is not conserved due to chiral anomaly effect and $\mathcal{C}$ denotes the anomaly coefficient. $\xi,\xi_5,\xi_B,\xi_{5B}$-terms are chiral anomaly-induced and correspond to non-dissipative transport phenomena, particularly making zero entropy production. In contrast, $\svq,\saq$, $\tilde{\sigma}_e,\tilde{\sigma}_{5e}$ are dissipative transport coefficients and have to be determined by the underlying microscopic theory and will be the focus of present work.


We would like to point out that studies in references \cite{Huang:2013iia,Jiang:2014ura,Pu:2014cwa,Pu:2014fva} were carried out under the {\it decoupling limit} (or {\it probe limit}), where the vector and axial currents were taken as decoupled from the stress-energy tensor of the fluid. While the decoupling limit allows deriving a simple Kubo formula for CESE conductivity, it does result in loss of some important physical contents, such as a pole in DC electrical conductivity due to spatial translation invariance \cite{Horowitz:2012ky, Blake:2013bqa, Davison:2015bea}. In the probe limit, the authors of \cite{Pu:2014cwa, Pu:2014fva} found a nonzero CESE conductivity for the holographic QCD model, which consists of probe $D8/\overline{D}8$-branes in the background geometry of $N_c$ stacks of $D4$-brane. A crucial point of \cite{Pu:2014cwa, Pu:2014fva} in realizing CESE is about the {\it explicit interaction} between the vector and axial gauge fields in the bulk, which originates from the nonlinear Dirac-Born-Infeld (DBI) action on the world-volumes of $D8/\overline{D}8$. Indeed, it was found that the CESE conductivity does vanish in a simple holographic setup with canonical $U(1)_V\times U(1)_A$ Maxwell fields {\it probing} the fixed Schwarzschild-AdS$_5$ background \cite{Gynther:2010ed,Bu:2016oba, Bu:2016vum}.

In this work, we reconsider the holographic $U(1)_V\times U(1)_A$ model and take into account the gravitational back-reaction effect on the gauge sectors of the bulk theory. While the anomaly coefficient $\mathcal{C}$ is fixed by the microscopic theory, we will take it as a free parameter and turn it off for the present study. Equivalently, the bulk Chern-Simons actions will be neglected in the holographic setup of \cite{Gynther:2010ed,Bu:2016oba, Bu:2016vum}. Note that there is no explicit interaction between $U(1)_V$ and $U(1)_A$ gauge fields in the bulk action. However, as will be clear later, the gravitational back-reaction will result in a coupling between perturbations of $U(1)_V$ and $U(1)_A$ bulk fields. This is the mechanism of generating CESE in such a simple holographic model.
In the next subsection, we will summarize the results and make the comparison with previous works. The remaining sections supplement all calculation details.

\subsection{Summary of the results} \label{summary}

Except for those chiral anomaly-induced terms, we will re-derive the first order constitutive relations \eqref{stress-cov},\eqref{covariantJm},\eqref{covariantJm5} via fluid/gravity correspondence \cite{Bhattacharyya:2008jc}. For a specific holographic model to be introduced in Section \ref{Model}, we analytically compute all dissipative transport coefficients. The calculation details will be presented in Section \ref{FluidGravity}. The shear and bulk viscosities in \eqref{pimunu} take universal values of holographic conformal fluids (holographically described by two-derivative Einstein gravity),
\begin{equation}
\eta=\frac{1}{4\pi}s,\qquad \qquad   \zeta=0,
\end{equation}
where $s$ is the entropy density of the dual fluid.

The dissipative transport coefficients in \eqref{covariantJm} and \eqref{covariantJm5}  could be presented in several different forms. In the first form, they are
\begin{align} \label{sigma=sigmaQ+sigma0}
\svq = \frac{{\sQ}\mv^2+  \sigma_0\ma^2}{\mv^2+\ma^2}, \qquad
\saq =\staq =  \mv \ma \frac{{\sQ}-\sigma_0}{\mv^2+\ma^2},\qquad
{\stq} = \frac{{\sQ} \ma^2+ \sigma_0 \mv^2}{\mv^2+\ma^2},
\end{align}
where $\sigma_Q$ and $\sigma_0$ are given by
\begin{align} \label{sigmaQ0}
{\sQ} \equiv  \frac{ \sigma_0  \,  (Ts)^2\,  }{\(Ts+{\mv}{\rn}+{\ma}{\rn}_5\)^2}=\frac{ \sigma_0 \,   (Ts)^2\,  }{\(\E+\P\)^2} ,\qquad
\sigma_0 \equiv \frac{\pi T}{2} \left(1+\sqrt{1+\frac{2}{3}\frac{\mu^2+\mu_5^2}{\pi^2T^2}}\right).
\end{align}
It is important to stress that $\svq,\saq$ and $\tilde{\sigma}_e,\tilde{\sigma}_{5e}$ are {\it intrinsic} conductivities, which are usually referred to as ``quantum critical'' \cite{Hartnoll:2007ih} or ``incoherent'' \cite{Hartnoll:2014lpa,Davison:2015taa} conductivities in holographic framework. We have reserved the notation $\sigma_Q$ to the intrinsic electrical conductivity for the single charge case, since $\svq=\sigma_Q$ once $\rho_5=0$ or $\mu_5=0$. In the probe limit where $\mu=\mu_5=0$, we have $\svq=\tilde{\sigma}_e=\sigma_0=\pi T$, but $\saq,\tilde{\sigma}_{5e}$ vanish.

In \eqref{sigma=sigmaQ+sigma0}, the relation $\saq=\staq$ originates from Onsager reciprocal relation for a time-reversal symmetric system. Additionally, there are mirror symmetric relations $\svq \leftrightarrow \stq$ and $\saq \leftrightarrow \staq$ under the exchange $\mu \leftrightarrow \mu_5$, which are specific to our holographic model \footnote{We thank Yan Liu for stimulating discussions on this issue.}. Thanks to these symmetric relations, in the discussions below we will focus on $\svq,\saq$. As will be clear in Section \ref{LinearAnalysis}, these symmetric relations for DC transport coefficients are extendable to associated alternating current (AC) conductivities in \eqref{conductivity matrix}. While the second law of thermodynamics, i.e., non-negativeness of divergence for the entropy current, could not fix the values of dissipative transport coefficients, it does set constraints for them \cite{Jiang:2014ura, Pu:2014cwa},
\begin{align}\label{constraints}
\svq\ge 0,\qquad \stq \ge 0, \qquad \svq \stq \ge \saq \staq,
\end{align}
which are obviously satisfied by our results \eqref{sigma=sigmaQ+sigma0}\eqref{sigmaQ0}.

From the dual fluid's point of view, it is more natural to re-express $\svq,\stq,\saq, \staq$ in terms of chemical potentials and temperature. As a result, \eqref{sigma=sigmaQ+sigma0} are turned into
\begin{align} \label{sigma_mumu5T}
\svq =\pi T \frac{2\Lc+(3\Lc-1)\bar{\mu}_5^2}{2(3\Lc-2)^2}\equiv T   \chiV, \qquad  
\saq =- \pi T \bar{\mu} \bar{\mu}_5 \frac{3\Lc-1}{2(3\Lc-2)^2}\equiv   T \bar{\mu}\bar{\mu}_5 \chiA, 
\end{align}
where
\begin{equation} \label{gamma}
\gamma(\bar{\mu},\bar{\mu}_5)=\frac{1}{2}\left(1+\sqrt{1+\frac{2}{3}(\bar{\mu}^2 +\bar{\mu}_5^2)} \right), \quad \qquad \bar{\mu}=\frac{\mu}{\pi T}, \qquad \bar{\mu}_5=\frac{\mu_5}{\pi T}.
\end{equation}
In \eqref{sigma_mumu5T} we defined $\chiA$ to measure deviation of the CESE conductivity $\saq$ from the conjectured universal factor $ T \bar{\mu} \bar{\mu}_5$ \cite{Huang:2013iia,Jiang:2014ura}. We also defined $\chiV$ to measure deviation of the Ohmic conductivity $\svq$ from its probe limit. In the high temperature (small chemical potential) limit or low temperature (large chemical potential) limit,  the conductivities \eqref{sigma_mumu5T} behave as,
\begin{equation}
\begin{split} \label{highT}
\svq&\,\simeq  \pi T, \qquad  \qquad\qquad ~~
\saq\,\simeq -\pi T \bar{\mu}\bar{\mu}_5,\qquad\qquad~~
\bar{\mu}, \bar{\mu}_5\ll1,\\
\svq&\,\simeq  \frac{\pi T \bar{\mu}_5^2} {\sqrt{6(\bar{\mu}^2+\bar{\mu}_5^2)}}, \qquad
\saq\, \simeq -\frac{\pi T \bar{\mu}\bar{\mu}_5} {\sqrt{6(\bar{\mu}^2+\bar{\mu}_5^2)}},\qquad  \bar{\mu}, \bar{\mu}_5\gg 1.
\end{split}
\end{equation}
Obviously, only in the high temperature regime where $\bar{\mu}, \bar{\mu}_5\ll1$ the CESE conductivity $\saq$ shows universal behavior as conjectured in \cite{Huang:2013iia,Jiang:2014ura}.

\begin{figure}[h]
\centering
\includegraphics[scale=0.53]{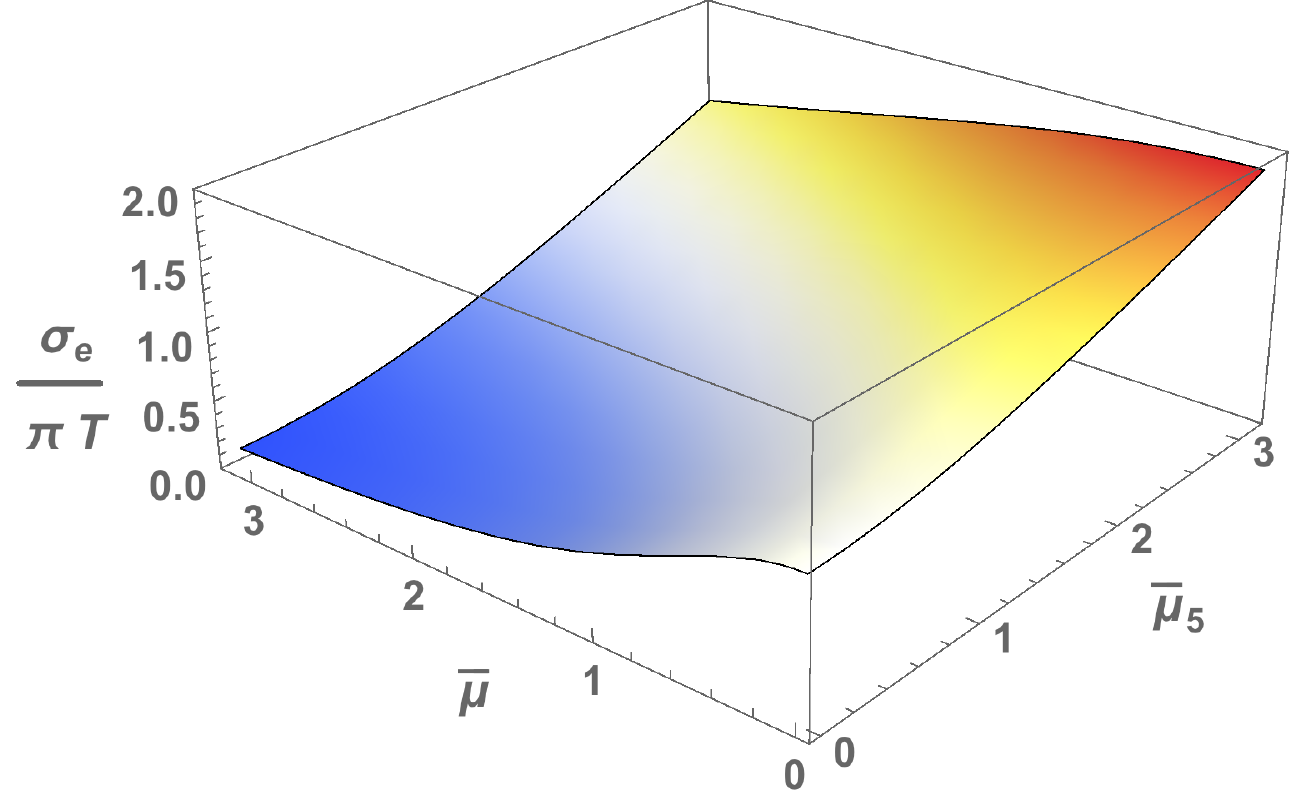}\qquad~~
\includegraphics[scale=0.53]{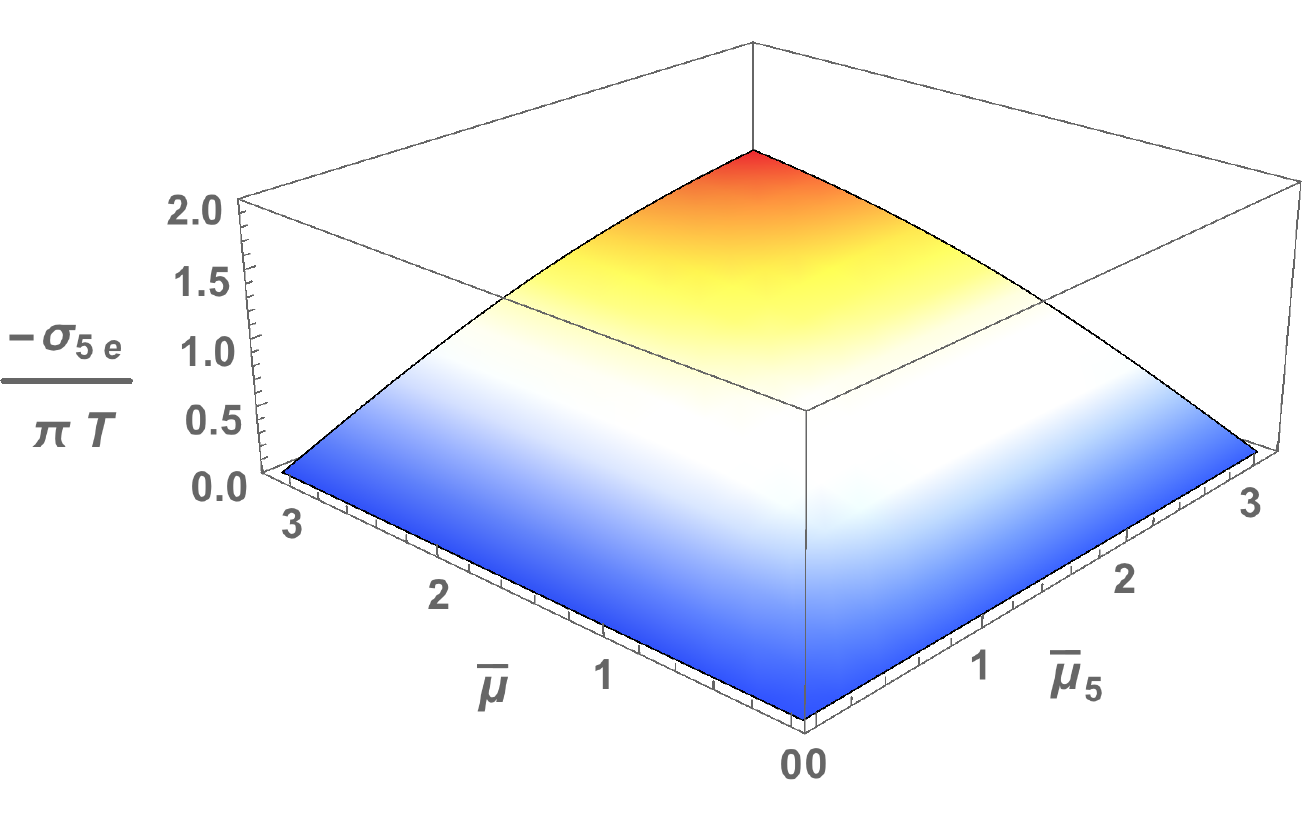}
\caption{The conductivities ${\svq}/(\pi T)$ and $-\saq/(\pi T)$ as functions of $\bar{\mu}$ and $\bar{\mu}_5$.}
\label{plot sigmae+sigma5e}
\end{figure}

\begin{figure}[h]
\centering\qquad
\includegraphics[scale=0.42]{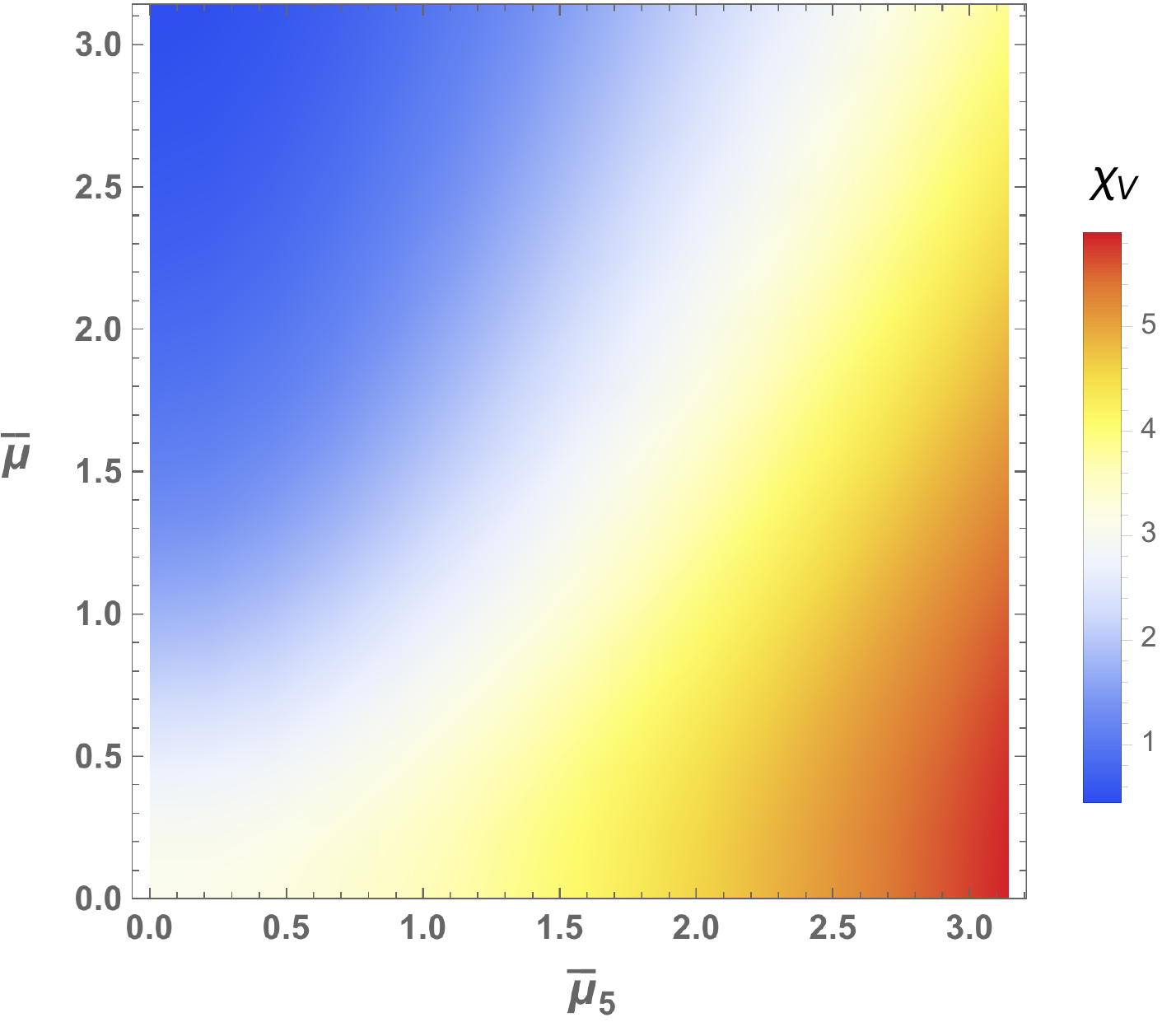}\qquad\qquad
\includegraphics[scale=0.42]{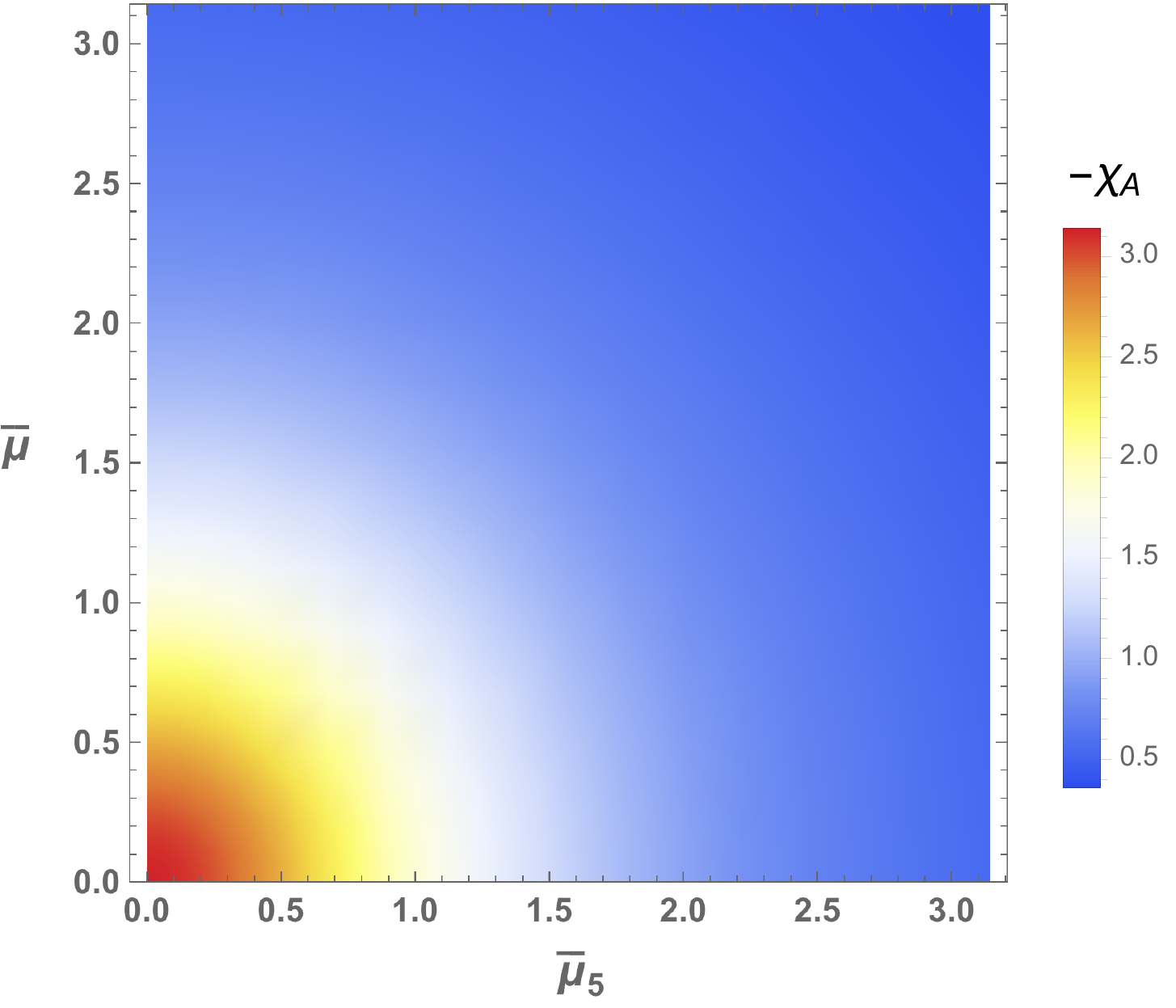}
\caption{Density plots for $\chiV$ and $\chiA$ in \eqref{sigma_mumu5T} as functions of $\bar{\mu}$ and $\bar{\mu}_5$.}
\label{chie}
\end{figure}

For illustration, in Figure \ref{plot sigmae+sigma5e} we plot ${\svq}/(\pi T)$ and $-\saq/(\pi T)$ as functions of dimensionless chemical potentials $\bar \mu$ and $\bar{\mu}_5$. As seen from Figure \ref{plot sigmae+sigma5e}, the axial chemical potential $\bar{\mu}_5$ has the effect of enhancing ${\svq}$ while the vector chemical potential $\bar{\mu}$ diminishes it. On the other hand, the CESE conductivity $\saq$ depends on both $\bar{\mu}$ and $\bar{\mu}_5$ non-trivially. Particularly, the deviation factor $\chiA$ is a monotonic decaying function of $\bar{\mu}$ and $\bar{\mu}_5$. In the high-temperature regime where $\bar{\mu},\bar{\mu}_5\ll1$, the factor $\chiA$ could be thought of as unity, which is obviously violated in the lower temperature limit. In Figure \ref{chie} we show density plots for the deviation factors $\chiV$ and $\chiA$.


Below we compare our results with those obtained within various other models \cite{Huang:2013iia,Jiang:2014ura,Pu:2014cwa,Pu:2014fva}, see Table \ref{table1}. The first two calculations in Table \ref{table1} were based on perturbative thermal QED and QCD, respectively. To leading-log order in gauge couplings, the conductivities were computed using the Kubo formula in \cite{Huang:2013iia,Jiang:2014ura}. The rest calculations in Table \ref{table1} are based on specific holographic models: Sakai-Sugimoto (S-S) model in \cite{Pu:2014cwa} versus holographic $U(1)_V\times U(1)_A$ model of present work.
As shown in Table \ref{table1}, the results obtained from  S-S model show the behavior of the pre-factors in $\svq,\saq\propto N_c^2 g_{\text{YM}}^2 $ \cite{Pu:2014cwa},
which is quite different from those in weakly coupled regime \cite{Huang:2013iia,Jiang:2014ura}.
While in our case, the conductivities depend on the bulk gauge couplings $\qV,\qA$.
If we employ the top-down setups in \cite{Erlich:2005qh, Matsuo:2009xn}, they could be related to the parameters of boundary theory by $1/\qV^2 \sim 1/\qA^2 \propto N_c N_f/ (4 \pi^2) $, where $N_c$ and $N_f$ are the numbers of the colors and flavors of the dual theory.

Although these dissipative transport coefficients are model dependent, in the high temperature regime, they do share some universal features, such as the linear dependence in $T$ (for $\svq$) and $T\bar\mu \bar\mu_5$ (for $\saq$). In addition, the Ohmic conductivity $\svq$ has to be positive \cite{Huang:2013iia}, but the sign of CESE conductivity $\saq$ could not be fixed by the second law of thermodynamics \cite{Huang:2013iia}. Particularly, $\saq$ was found to be positive for the models in \cite{Huang:2013iia,Jiang:2014ura,Pu:2014cwa}, but it is shown to be negative from present study \eqref{sigma_mumu5T}. As shown in Table \ref{table1}, from weak to strong coupling, the conductivities $\svq,\saq$ show different dependence on the gauge coupling of boundary gauge theory in various models.


\begin{table}[h]

\caption{Ohmic and CESE Conductivities in The High Temperature Regime in Various Models}
\label{table1}
\begin{tabular}{|c|c|c|c|c|}%
\hline  %
Pre-factors &QED Plasma \cite{Huang:2013iia}
& QGP ($u,d$)   \cite{Jiang:2014ura}
& S-S Model \cite{Pu:2014cwa}
&  $U(1)_V\times U(1)_A$\\
\hline  %
$\chiV\equiv \sigma_e/T$
&$ \frac{15.7}{e^3\ln(1/e)}   $
&$ 13.0 \frac{\text{Tr}_f Q_e Q_V  }{g_c^4\ln(1/g_c)}    $
&$0.025 \frac{8 N_c^2 g_{_{\text{YM}}}^2 }{81 }   $
& $  {\pi}/{\qV^2}$ ~   [(\ref{highT})]  \\
\hline %
$\chiA\equiv \sigma_{5e}/(T \bar{\mu}\bar{\mu}_5)  $
&$ \frac{20.5}{e^3\ln(1/e)} $  
&$14.5\frac{\text{Tr}_f Q_e Q_A  }{g_c^4\ln(1/g_c)}  $
&$0.002 \frac{8 N_c^2 g_{_{\text{YM}}}^2 }{81 }   $  
& $-  {\pi}/{\qA^2}$    [(\ref{highT})] \\
\hline
\end{tabular}
\vspace{5pt}

{\bf Note}: {
In this table, $e,\  g_c,\ g_{\rm YM}$ are the gauge couplings of QED, QCD, the dual $SU(N_c)$ gauge theory, respectively.
For the calculations in QGP with two light quarks  $(u,d)$, $Q_e=\text{Diag}(2/3, -1/3)$, $Q_V$ and $Q_A$ are the vector and axial charge matrices in flavor space. 
For the  S-S model, the results above are read off from relevant numerical plots around $T\simeq 0.2$GeV \cite{Pu:2014cwa}.
Actually, in the high-temperature regime where the chemical potentials are suppressed, $\sigma_e\propto T^2/m_{KK}$ and $m_{KK}$ is the Kaluza-Klein mass in the S-S model \cite{Pu:2014fva,Muller:2015maa}.
For the results of present study, we have restored the bulk gauge couplings $\qV^2$ and $\qA^2$, which are related to parameters of boundary theory by $1/\qV^2 \sim 1/\qA^2 \propto   N_c N_f/ (4 \pi^2) $ \cite{Erlich:2005qh, Matsuo:2009xn}.
}
\end{table}

From Table \ref{table1},  in the high temperature regime that $T \gg \mu, \mu_5 $, the conductivities in holographic $U(1)_A\times U(1)_V$ model share the same temperature scalings as those in the QED plasma or QGP.
The Ohmic conductivity depends linearly on the temperature,  $\svq\propto T$, while the CESE conductivity depends inversely on the temperature, $\saq\propto \frac{\mu \mu_5}{T}$.
It will become more evident from dimensional analysis as implemented in \cite{Jiang:2014ura} within relativistic kinetic theory.
In the high temperature regime, the conductivities can be expanded in terms of the ratios $\(\frac{\mu}{T}\)$ and $\(\frac{\mu_5}{T}\)$. To quadratic order in the ratios, the conductivities are constrained to be  $\svq \simeq f(T) \[1+ d_{20} \(\frac{\mu}{T}\)^2+ d_{02} \(\frac{\mu_5}{T}\)^2\]$ and $\saq \simeq f(T) d_{11} \(\frac{\mu}{T}\)  \(\frac{\mu_5}{T}\) $, where $f(T)$ is a function of the temperature and $d_{02}, d_{20}, d_{11}$ are constants \cite{Jiang:2014ura}. We know that the conductivities have the same dimension as the temperature or chemical potentials, $[\sigma_5]= [\sigma_{5e}] = [T]= [\mu]= [\mu_5]$.  So, if there are no extra dimensional physical quantities in the model, it is valid to conclude that $f(T)\simeq T$, which also supports the general scaling behaviors as summarized in Table \ref{table1}.

On the universal scaling behaviors in the high temperature regime, for a physical explanation, it is nature to assume that the interaction between the charges in the fluid will become weaker in the higher temperature.  The Ohmic conductivity $\svq$ describes the response of the charged vector current to the external electrical field, $\svq \simeq \vec{J} /\vec{E}$. When the temperature is increased, $\svq$ will be enhanced because the moving of the charges in the fluid will become easier.
However, the CESE conductivity $\saq$ measures the response of the axial current to the external electrical field, $\saq \simeq  \vec{J}_5 /\vec{E}$. When temperature is increased, the interaction between the axial current and electrical field will become weaker and $\saq$ will decrease due to the additional interaction factor $\(\frac{\bar{\mu}}{T}\)\(\frac{\bar{\mu}_5}{T}\)$.

While the CESE is a non-anomalous transport phenomenon, it may induce phenomenological consequences in heavy-ion collisions, namely the net charge distribution and correlation patterns in Cu+Au collisions as discussed in \cite{Huang:2013iia}. Admittedly, this should be a mixture due to CESE, CME, and CSE. However, CESE, along with other robust anomalous transport phenomena, is masked by various backgrounds in heavy-ion collisions,
making it very difficult to pin down, not even to explore its properties. On the other hand, the exotic topological states of metal, such as Dirac and Weyl semi-metals, provide an experimental playground to study potential observable effects of CESE and other anomalous transports in a controllable way.
See \cite{Gorbar:2017lnp,Gorbar:2017vph, Gorbar:2018vuh, Gorbar:2018nmg} for recent progress on this topic, as well as the relevant investigations from holographic models \cite{Landsteiner:2015lsa, Landsteiner:2015pdh, Landsteiner:2016stv, Landsteiner:2014vua, Sun:2016gpy, Seo:2016vks, Rogatko:2017svr, Rogatko:2017tae, Ammon:2018wzb}.

Below we would like to rewrite the currents \eqref{covariantJm} and \eqref{covariantJm5} in a linear response form, in which the electric field $E_i$ and thermal gradient $\nabla_i T$ are taken as external sources. In other words, the fluid velocity $u_i$ will be eliminated using the current conservation law in \eqref{hydro eqs}. Here, the chemical potentials $\mu,\mu_5$ will be taken as constant. Consequently,
\begin{align}
u^i=u_i=\frac{1}{i\omega} \Big( s \nabla_i T- \frac{\rn}{\E+\P} {\Em}_i \Big),
\end{align}
where $\partial_t\to -i\omega$ is used. Then, the currents \eqref{covariantJm} and \eqref{covariantJm5} turn into
\begin{align}
{J}^t &={\rn},\qquad  \qquad  ~~
{J}_i=\svo E_i- \avo  {\nabla_i T}, \label{Jmu_linear}\\
{J}^t_5&={\rn}_5,\qquad   \qquad
{J}_{5i}= \sao  E_i-  \aao  {\nabla_i T}, \label{Jmu5_linear}
\end{align}
where
\begin{align}
\svo&=\frac{i}{\omega} \frac{{\rn}^2}{\E+\P}+{\svq},\qquad  \qquad \qquad  ~~
\sao=\frac{i}{\omega} \frac{{\rn}{\rn}_5}{\E+\P}+ {\saq}, \label{sigmaVo}\\
\avo{T}& =\frac{i}{\omega} {\rn} - \( \mu\, \svo + \mu_5\,\sao\),\qquad
\aao{T}=\frac{i}{\omega} {\rn}_5 - \( \mu_5\, \sto + \mu\,\sao\).\label{sigmaAo}
\end{align}
In \eqref{sigmaAo}, $\sto$ is related to $\stq$ by
\begin{align} \label{sigmat_omega}
\sto\equiv\svo \big |_{{\rn} \leftrightarrow {\rha}}=\frac{i}{\omega} \frac{{\rn}_5^2}{\E+\P}+\stq,
\end{align}
where $\stq$ is presented in \eqref{sigma=sigmaQ+sigma0}.
Physically, $\svo$ is the low-frequency limit of the Ohmic electrical conductivity, and $\sao$ measures the chiral electrical separation effect. Obviously, in the probe limit, $\svo$ and $\sao$ are dominated by their intrinsic parts ${\svq}$ and ${\saq}$.
$\avo$ and $\aao$ are the thermoelectric conductivities for vector and axial currents. The heat current is
\begin{align} \label{heat_linear}
\Q^i\equiv {T^{ti}} -\mu {J}^i- \mu_5 J_5^i =\, \bavo T E_i - \bkvo \nabla_i T,
\end{align}
where the transport coefficients are
\begin{align} \label{heat-conduc-hydro}
\bavo =\avo,\quad
\bkvo T=  \frac{i}{\omega}  \(\E+\P-2{\rn}\mv -2{\rn}_5\ma\)  +
\( \mu^2\, \svo+\mu_5^2\, \sto +2\mu \mu_5 \sao\).
\end{align}
$\bkvo$ is the  low-frequency limit of thermal conductivity, and it is fully determined by the intrinsic conductivities ${\svq},\saq$ in \eqref{sigma=sigmaQ+sigma0}.
Here we would like to emphasize that $\svo$, $\sao$, $\sto$, $\avo$, $\aao$, $\bavo$, $\bkvo$ are different from the intrinsic conductivities $\svq,\saq, \tilde{\sigma}_e, \tilde{\sigma}_{5e}$: while the former could be directly read off from associated Kubo formulas, the latter are useful in parameterizing the hydrodynamic constitutive relations \eqref{covariantJm}\eqref{covariantJm5}. In the probe limit, the differences between them are absent.

As the second study of the present work, we compute frequency-dependent conductivities. The external sources are vector and axial external fields $E_i,E_i^5$ and temperature gradient $\nabla_i T$, which depend on time via a plane waveform. As a result, the low-frequency conductivities $\svo$, $\sao$, $\sto$, $\avo$, $\aao$, $\bavo$, $\bkvo$ are generalized to frequency-dependent AC conductivities, see \eqref{conductivity matrix}. First, we analytically evaluate the low-frequency limits of all the AC conductivities, demonstrating agreement with \eqref{sigmaVo},\eqref{sigmaAo},\eqref{sigmat_omega},\eqref{heat-conduc-hydro}. For the general value of frequency, we numerically compute all AC conductivities, see plots in Section \ref{LinearAnalysis}.

Note the appearance of $i/\omega$ pieces in the physical conductivities $\svo$ and $\sao$. Via Kramers-Kronig relations \cite{Hartnoll:2008vx,Hartnoll:2008kx}, this $i/\omega$ immediately implies the existence of a $\delta(\omega)$ in real parts of $\svo$ and $\sao$, as is required by translation invariance.
However, in holographic models with spatial translational symmetry, it is hard to reveal the delta peak in DC conductivity analytically. Once translational symmetry is broken, the DC conductivities become finite with the delta peak removed. Within the relaxation time approximation, the momentum dissipation effect would result in the replacement
\begin{equation}
\frac{i}{\omega}=\frac{1}{-i\omega}~~{\rightarrow}~~ \frac{1}{1/\tau-i\omega},
\end{equation}
where $\tau$ corresponds to momentum relaxation time. Now all physical conductivities become finite and, in particular, they are split into two parts: the ``coherent'' pieces (related to the momentum dissipation) and the ``inherent'' ones (the universal pieces). Admittedly, a more rigorous treatment of momentum dissipation along the line \cite{Horowitz:2012ky, Blake:2013bqa, Davison:2015bea, Blake:2015epa, Blake:2015hxa} would be useful in clarifying physical meanings of these transport coefficients, and we will address this elsewhere.

The remaining sections are structured as follows. In Section \ref{Model}, we present the holographic model. In Section \ref{FluidGravity}, with anomalous terms neglected, we re-derive the first-order constitutive relations (\ref{stress-cov}) (\ref{covariantJm}) (\ref{covariantJm5}) using the fluid/gravity correspondence, and analytically compute all dissipative transport coefficients.
In Section \ref{LinearAnalysis}, we obtain AC conductivities through linear response analysis. Section \ref{Conclusion} contains the conclusion and discussions. Two appendices \ref{appF} and \ref{appL} provide further calculation details.

{\bf Notation Conventions}: we use the upper-case Latin letters $M,N,\cdots$ to denote the $(4+1)$-dimensional bulk coordinates, the lower-case Greek letters $\mu,\nu,\cdots$ to denote the $(3+1)$-dimensional boundary directions, while $i,j,\cdots$ will be used for spatial directions on the boundary.

\section{Holographic Model for Fluid with $U(1)_V\times U(1)_A$ Currents}
\label{Model}

We consider $(4+1)$-dimensional Einstein gravity with a negative cosmological constant $\Lambda=-6/L^2$ in the bulk,  which is coupled to $U(1)_V\times U(1)_A$ gauge fields (see, e.g., \cite{Gynther:2010ed,Bu:2016oba}).
The total bulk action is
\begin{align}  \label{bulkaction}
{\S_\M}&=\frac{1}{16\pi G_5} \int_\M \d^5x \sqrt{-g}\left(R-2\Lambda \right)+\S_F + \S_{\textrm{ct}} + \S_{\K},
\end{align}
where
\begin{align}\label{GaugeCoupling}
\S_F&= -\int_\M {\d}^5x \sqrt{-g}\, \bigg( \frac{1}{4{\qV^2}} \Fv_{MN}\Fv^{MN} +  \frac{1}{4{\qA^2}}\Fa_{MN}\Fa^{MN}\bigg).
 \end{align}
The field strengths of the bulk $U(1)$ gauge fields are defined as
$\Fv_{MN} \equiv \partial_M {\Av}_N- \partial_N {\Av}_M$ and
$\Fa_{MN} \equiv \partial_M {\Aa}_N- \partial_N {\Aa}_M$. 
In our notations, ${\Av}_M$ and ${\Aa}_M$ denote the vector and axial bulk gauge fields, which are dual to vector and axial currents $J^\mu,J^\mu_5$ of the boundary conformal field theory (CFT), respectively. The Gibbons-Hawking-York term $\S_{\K}$ is
\begin{align}
\S_{\K}&=\frac{1}{8 \pi G_5} \int_{\p\M} {\d}^4x \sqrt{-\gamma} \,\K, \label{GHY}
\end{align}
where $\K=\gamma^{\mu\nu} \K_{\mu\nu}$. $\gamma_{\mu\nu}$ is the induced metric on the boundary  hypersurface $\Sigma$ defined by the equation $r=r_c$, and $\K_{\mu\nu}$ is the extrinsic curvature tensor on $\Sigma$,
\begin{align}\label{extrinsic}
\K_{\mu\nu}
=\frac{1}{2}\mathcal{L}_n \gamma_{\mu\nu}\equiv\frac{1}{2} \left(n^M \partial_{M} \gamma_{\mu\nu}+\gamma_{\mu N} \partial_\nu n^N+ \gamma_{\nu N} \partial_\mu n^N\right),
\end{align}
where $\mathcal{L}_n$ is the Lie derivative along the unit normal vector $n_M$ of the hypersurface $\Sigma$:
\begin{equation}
n_M=\frac{\partial_{M} r}{\sqrt{g^{AB}\partial_{A} r \partial_B r}}.
\end{equation}

The counter-term action $\S_{\rm ct}$ is \cite{Henningson:1998gx, Balasubramanian:1999re, Taylor:2000xw,deHaro:2000vlm,Matsuo:2009xn,Sahoo:2010sp}
\begin{align}
\S_{\textrm{ct}}& = -\frac{1}{16\pi G_5} \int_{\p\M} {\d}^4x \sqrt{-\gamma}\left(  \frac{1}{2} \R + 6\right) +  \frac{ \log{r_c} }{4} \int_{\p\M} {\d}^4x \sqrt{-\gamma}\left(\frac{1}{ {\qV^2}} {\Fv}_{\mu\nu} {\Fv}^{\mu\nu}
+ \frac{1}{ \qA^2}{\Fa}_{\mu\nu} {\Fa}^{\mu\nu} \right), \label{counter}
\end{align}
where $\R$ is the Ricci scalar of the induced metric $\gamma_{\mu\nu}$. Note minimal subtraction scheme has been utilized in writing down the counter-terms for bulk Maxwell fields.  ${\Fv}_{\mu\nu}$ and ${\Fa}_{\mu\nu}$ in (\ref{counter}) are the projections of bulk field strengths $\Fv_{MN}$ and $\Fa_{MN}$ onto the hypersurface $\Sigma$.

We would like to stress once again that the possible Chern-Simons terms $\Aa\wedge \Fv \wedge \Fv$ and $\Aa\wedge \Fa \wedge \Fa$ have been ignored in (\ref{bulkaction}), which amounts to switching off the chiral anomaly effect in the dual boundary theory \cite{Gynther:2010ed,Bu:2016oba}.
Indeed, in order for the chiral anomaly to take effect, the dual plasma should either be exposed to a magnetic environment or rotate. Thus, with an electric field as the only external source, all the anomalous transports in \eqref{covariantJm}\eqref{covariantJm5} vanish accidentally.

Variation of total bulk action ${\S_\M}$ in \eqref{bulkaction} with respect to bulk metric $g_{MN}$ gives rise to the Einstein equation,
\begin{align}
\Ew_{MN} & \equiv  \,R_{MN}-\frac{1}{2} R\, g_{MN} -6g_{MN} -\left(T_{MN}^{\rm bulk}+ {\tT}_{MN}^{\rm bulk}\right)=0, \label{Einstein}
\end{align}
where
\begin{align}
T_{MN}^{\rm bulk}&=  \frac{\Up}{2} \Big( {\Fv}_{AM} {\Fv}^A_{~~N}-\frac{1}{4} g_{MN} {\Fv}^2 \Big),\qquad
\Up \equiv \frac{16\pi G_5}{{\qV^2}}, \nn\\
{\tT}_{MN}^{\rm bulk}&= \frac{{\tUp}}{2}\Big( {\Fa}_{AM} {\Fa}^A_{~~N}-\frac{1}{4} g_{MN} {\Fa}^2\Big),\qquad
{\tUp}\equiv \frac{16\pi G_5} {\qA^2}. \nn
\end{align}
The bulk stress-energy tensor has been denoted as $T_{MN}^{\rm bulk}$ and $\tT_{MN}^{\rm bulk}$, which should not be confused with that of the dual boundary theory. $\Up$ and ${\tUp}$ measure the strength of back-reaction of bulk gauge fields on the bulk geometry.
The Maxwell equations for $\Av_M$ and $\Aa_M$ are,
\begin{align}
\Ev^N&\equiv\nabla_{M}{\Fv}^{MN}=\frac{1}{\sqrt{-g}}\partial_{M}\big(\sqrt{-g}{\Fv}^{MN}\big)=0, \label{MaxwellV}\\
\Ea^N&\equiv\nabla_{M}{\Fa}^{MN}=\frac{1}{\sqrt{-g}}\partial_{M}\big(\sqrt{-g}{\Fa}^{MN}\big)=0. \label{MaxwellA}
\end{align}

According to Anti-de Sitter/Conformal Field Theory (AdS/CFT) correspondence \cite{Maldacena:1997re, Gubser:1998bc, Witten:1998qj}, the expectation values of stress-energy tensor and currents on the boundary theory are defined as
\begin{equation} \label{T+J+J5 def}
T_{\mu\nu}\equiv \lim_{r\to\infty} \frac{-2r^2}{\sqrt{-\gamma}} \frac{\delta {\S_\M}}{\delta \gamma^{\mu\nu}} ,\qquad \quad
J^\mu\equiv \lim_{r\to\infty}  \frac{\delta {\S_\M}}{\delta {\Av}_\mu}, \qquad \quad
J^\mu_5\equiv\lim_{r\to\infty}  \frac{\delta {\S_\M}}{\delta {\Aa}_\mu}.
\end{equation}
In terms of bulk fields, (\ref{T+J+J5 def}) turns into
\begin{align}
T_{\mu\nu}&=- \frac{1}{8\pi G_5}\lim_{r\to\infty} r^2 \Big[  \big(\K_{\mu\nu} - \K \gamma_{\mu\nu} + 3 \gamma_{\mu\nu} -\frac{1}{2}\mathcal{G}_{\mu\nu} \big)+\T^{F}_{\mu\nu}\Big], \label{Tmunu} \\
J^\mu &=-\frac{1}{{\qV^2}}\lim_{r\to\infty} r^2 \big(n_M {\Fv}^{M\mu}+  D_\nu {\Fv}^{\nu\mu}\log r \big), \label{Jmu}\\
J^\mu_5 &=-\frac{1}{\qA^2}\lim_{r\to\infty}  r^2 \big( n_M {\Fa}^{M\mu} + D_\nu {\Fa}^{\nu\mu} \log r\big), \label{J5mu}
\end{align}
where the counter-term
\begin{align}\label{stressFct}
\T^{F}_{\mu\nu}\equiv \frac{\log r}{4{\qV^2}}\Big({\Fv}_{\alpha\mu} {\Fv}^\alpha_{~\nu} -\frac{1}{4}\gamma_{\mu\nu} {\Fv}^{2}\Big)+\frac{\log r}{4\qA^2}\Big({\Fa}_{\alpha\mu} {\Fa}^\alpha_{~\nu} -\frac{1}{4}\gamma_{\mu\nu} {\Fa}^{2}\Big),
 \end{align}
which will affect the study of Section \ref{LinearAnalysis} beyond first-order transport coefficients. In (\ref{Tmunu})-(\ref{J5mu}), $\mathcal{G}_{\mu\nu} $ is the Einstein tensor associated with the induced metric $\gamma_{\mu\nu}$,
and $D$ is the covariant derivative operator compatible with $\gamma_{\mu\nu}$.

The bulk equations \eqref{Einstein}-\eqref{MaxwellA} can be classified into dynamical components and constraint ones. Moreover, the constraint components correspond to conservation laws for boundary stress-energy tensor and currents:
\begin{equation} \label{conservation1}
\begin{split}
&{\Ew}^r_{~\nu}=0 ~\Rightarrow~ \partial^\mu T_{\mu\nu}
=J^\alpha  {\Fv}^{\rme}_{\nu\alpha}+  J_5^\alpha {\Fa}^{\rme}_{\nu\alpha} ,\\
&{\Ev}^r=0 ~~\Rightarrow~ \partial_\mu J^\mu=0,\\
&{\Ea}^r=0 ~~\Rightarrow~ \partial_\mu J_5^\mu=0,
\end{split}
\end{equation}
where ${\Fv}^{\rme}_{\nu\alpha}$ and ${\Fa}^{\rme}_{\nu\alpha}$ are external vector and axial electromagnetic field strengths, respectively.


In what follows, we will work under the convention $16\pi G_5=1$ and $L=1$. The bulk gauge couplings $\qV, \qA$ will be absorbed into redefinitions of bulk gauge fields ${\Av}_M\to {\qV} {\Av}_M$ and ${\Aa}_M\to {\qA} {\Aa}_M$.
The boundary CFT in thermal equilibrium corresponds to a homogeneous solution of the bulk theory (\ref{bulkaction}). We assume the presence of finite vector and axial charge densities. Consequently, the homogeneous solution of the bulk theory is the AdS$_5$ black brane with two charges,
\begin{equation} \label{homogeneous}
\begin{split}
ds^2_{\eq}&=g_{MN}^{\eq}{\d}x^M{\d}x^N=2{\d}t{\d}r-r^2f(r){\d}t^2+r^2\delta_{ij} {\d}x^i {\d}x^j,\\
f(r)&\equiv 1-\frac{M}{r^4}+\frac{Q^2+{{\tQ}}^2}{r^6}=\frac{\left(r^2-\rh ^2\right) \left(r^2-r_-^2\right)
\left(r^2+\rh ^2+r_-^2\right)}{r^6},\\
{\Av}_{\eq} &=-\frac{\sqrt{3}Q}{r^2}{\d}t,\qquad\qquad {\Aa}_{\eq}=-\frac{\sqrt{3} {{\tQ}}}{r^2}{\d}t,
\end{split}
\end{equation}
where $M, Q,{{\tQ}}$ are constant parameters of the bulk theory.

In (\ref{homogeneous}),  $\rh $ is the largest root for $f(r)=0$, defining the location of event horizon of the two-charges AdS$_5$ black brane. The Hawking temperature, identified as the temperature of dual boundary field theory, is
\begin{equation} \label{Hawking temperature}
T_0=\frac{\partial_r\left(r^2f(r)\right)}{4\pi}\Big|_{r=\rh }= \frac{\rh }{\pi} \bigg(1-\frac{Q^2+{{\tQ}}^2}{2\rh ^6}\bigg),
\end{equation}
which should be non-negative, setting constraints on the combination $(Q^2+{{\tQ}}^2)$. At the horizon with a constant time section, the line element degenerates into
$ds^2|_{\textrm{Horizon}}=\rh ^2\delta_{ij}{\d}x^i{\d}x^j$, with $i,j=1,2,3$.
Thus, the entropy density of the black brane (\ref{homogeneous}), which will be identified as that of the dual CFT, turns out to be
\begin{equation}\label{entropy density}
s_0=\frac{\text{Area}(\rh )}{4G_5}=\frac{\rh ^3}{4G_5}=4\pi \rh ^3,
\end{equation}
where in the second equality we made use of the normalization convention $16\pi G_5=1$. The subscript ``$_0$'' in (\ref{Hawking temperature}) (\ref{entropy density}) is to emphasize that they are constant, as compared to temperature field $T(x^\alpha)$ in Section \ref{FluidGravity}.

Finally, via \eqref{Tmunu}-\eqref{J5mu}, the dual stress-energy tensor and currents of the boundary theory are
\begin{equation} \label{eq TJ}
\begin{split}
T^{\mu\nu}_{\eq}= 3M \, \delta_t^\mu\delta_t^\nu+M \, \delta_i^\mu\delta_j^\nu , \qquad
J^\mu_{\eq} =  2\sqrt{3}Q\,  \delta_t^\mu,  \qquad
J^\mu_{5\eq}  = 2\sqrt{3}{\tQ}\, \delta_t^\mu ,
\end{split}
\end{equation}
where the subscript ``$_{\eq}$'' is to mark that these quantities are associated to a state in thermal equilibrium. If we make the following identifications,
\begin{equation} \label{eq energy-charge}
\E= 3M, \qquad  \P=M, \qquad  \rho= 2\sqrt{3}Q, \qquad  \rho_5= 2\sqrt{3}{\tQ},
\end{equation}
then (\ref{eq TJ}) are nothing but the non-derivative parts of the stress-energy tensor and currents  (\ref{stress-cov}) (\ref{covariantJm}) (\ref{covariantJm5}) in local rest frame where $u_\mu=(-1,0,0,0)$.
%
%
In the next two Sections \ref{FluidGravity} and \ref{LinearAnalysis} we will solve the bulk dynamics with equations of motion (\ref{Einstein}) (\ref{MaxwellV}) (\ref{MaxwellA}) under two complementary limits,
hydrodynamic limit versus linear approximation, generating viscous corrections to ideal fluid (\ref{eq TJ}).

\section{First order hydrodynamics from fluid/gravity correspondence} \label{FluidGravity}


In this section, we construct the first-order hydrodynamics dual to the bulk theory (\ref{bulkaction}) via the fluid/gravity correspondence \cite{Bhattacharyya:2008jc,Bhattacharyya:2008mz}.

\subsection{Set Up the Fluid/Gravity Calculations} \label{setup fluid/gravity}

In this subsection, we set up the stage for performing fluid/gravity calculations for the bulk theory of \eqref{bulkaction}.
Following the standard procedure of fluid/gravity correspondence \cite{Bhattacharyya:2008jc, Bhattacharyya:2008mz}, we make a Lorenz boost transformation for the static solution (\ref{homogeneous}) along the boundary coordinates
\begin{equation} \label{lorentz bs}
x^\mu\to {L^\mu}_\nu x^\nu,\qquad
{L^t}_\nu= - u_\nu,\qquad {L^i}_{\nu}=\big(-u^i,\delta^i_j+\frac{u^iu_j}{1-u_0}\big).
\end{equation}
${L^\mu}_\nu$ is the Lorentz boost matrix, which has been parameterized via a four-velocity $u_\mu$.
Note the four-velocity $u_\mu$ is a time-like unit vector obeying
$\eta^{\mu\nu}u_\mu u_\nu=-1$, which leads to $u_0=-\sqrt{1+\vec{u}^2}$.
After the transformation (\ref{lorentz bs}), the homogenous solution \eqref{homogeneous} turns into
\begin{equation}\label{rnboost1}
\begin{split}
&ds^2_{\bs} = - r^2 f(r)( u_\mu {\d}x^\mu )^2 - 2 u_\mu {\d}x^\mu dr + r^2(\eta_{\mu \nu} +u_\mu u_\nu) {\d}x^\mu {\d}x^\nu,\\
&{\Av}_{\bs} =   \frac{\sqrt 3 Q}{r^2} u_\mu {\d}x^\mu  +  {\Av}^{\rme}_\mu {\d}x^\mu,\qquad
{\Aa}_{\bs}=  \frac{\sqrt{3} {{\tQ}}}{r^2} u_\mu {\d}x^\mu,\\
&f(r)=  1- \frac{M}{r^4} +\frac{Q^2 +{{\tQ}}^2}{r^6},
\end{split}
\end{equation}
where a constant vector field ${\Av}^{\rme}_\mu$ is introduced for the later purpose of exposing the boundary theory to an external electric field environment. So long as the parameters $M,Q,{{\tQ}},{\Av}^{\rme}_\mu$ are constants, the boosted solution (\ref{rnboost1}) does solve the bulk equations of motion \eqref{Einstein}-\eqref{MaxwellA}.

One key procedure of fluid/gravity correspondence is to promote the constant parameters in (\ref{rnboost1}) to arbitrary functions of boundary coordinates \cite{Bhattacharyya:2008jc,Bhattacharyya:2008mz,Hur:2008tq,Son:2009tf},
\begin{equation} \label{promotion}
M\to M(x^\alpha),\quad Q\to Q(x^\alpha),\quad {{\tQ}}\to {{\tQ}}(x^\alpha),\quad
u_\mu\to u_\mu(x^\alpha), \quad {\Av}^{\rme}_\mu \to {\Av}^{\rme}_\mu(x^\alpha).
\end{equation}
Then, these nontrivial functions $M(x),Q(x),{{\tQ}}(x),u_\mu(x)$ and ${\Av}^{\rme}_\mu(x)$ are identified as the fluid-dynamical variables and external source of the dual field theory. However, after the promotion (\ref{promotion}) the boosted solution (\ref{rnboost1}) will not satisfy the bulk equations \eqref{Einstein}-\eqref{MaxwellA} any more. That is, the price of the promotion (\ref{promotion}) is that one has to add suitable corrections to the metric and gauge fields in the bulk theory so that the bulk equations \eqref{Einstein}-\eqref{MaxwellA} can be obeyed. For general functions $M(x),Q(x),{{\tQ}}(x),u_\mu(x)$ and ${\Av}^{\rme}_\mu(x)$, it is very difficult to work out these suitable corrections. The fluid/gravity correspondence shows that in the hydrodynamic limit, where the functions $M(x),Q(x),{\tQ}(x),u_\mu(x)$ and ${\Av}^{\rme}_\mu(x)$ vary rather slowly from point to point, the corrections can be systematically collected order-by-order within a boundary derivative expansion.


In the practical calculation, we do Taylor expansion for $M(x),Q(x),{{\tQ}}(x),u_\mu(x)$ and $\Av^{\rme}_\mu(x)$ around the point of origin $x^\mu=0$,
\begin{equation}
\begin{split}
M(x)&=M_0+{\ve} x^\alpha \partial_\alpha M{\0}+\o(\partial^2),\qquad
u_\mu(x)=(-1,{\ve} (x^\alpha\partial_\alpha u_i{\0}))+\o(\partial^2),\\
Q(x)&=Q_0+{\ve} x^\alpha \partial_\alpha Q{\0}+\o(\partial^2),\qquad~~
{{\tQ}}(x)={{\tQ}}_0+{\ve} x^\alpha \partial_\alpha {{\tQ}}{\0}+ \o(\partial^2),\\
{\Av}^{\rme}_\mu(x)&=0+{\ve} x^\alpha \partial_\alpha {\Av}^{\rme}_\mu{\0} +\o(\partial^2),
\end{split}
\end{equation}
where we have chosen the frame that $u_i{\0}=0$ and $\Av^{\rme}_\mu{\0}=0$ at the origin. Moreover, a formal parameter ${\ve}$ is introduced to count the number of derivatives in the expansion, and eventually will be set to unity. All calculations in this section will be accurate up to the first-order in the derivative expansion. Consequently, up to order $\mathcal{O}(\ve^1)$ the promoted metric and gauge fields become
\begin{align} \label{seed}
ds^2_{\seed}=& g_{MN}^{\seed} {\d}x^M {\d}x^N= \,2{\d}t{\d}r-r^2f_0(r){\d}t^2+r^2\delta_{ij}{\d}x^i {\d}x^j\nn\\
&\qquad\qquad\qquad - \ve \left\{r^2 f_1(r,x){\d}t^2 -2r^2[f_0(r)-1](x^\alpha\partial_\alpha u_i{\0}) {\d}t{\d}x^i+ 2(x^\alpha\partial_\alpha u_i{\0}) {\d}x^i dr \right\},\nn\\
{\Av}^{\seed}=&{\Av}^{\seed}_{M} {\d}x^M=- \frac{\sqrt{3}(Q_0+x^\alpha \partial_\alpha Q{\0})}{r^2}dt + \ve \frac{\sqrt{3}Q_0}{r^2} (x^\alpha\partial_\alpha u_i{\0}) {\d}x^i + x^\alpha \partial_\alpha {\Av}_\mu^{\rme}{\0}{\d}x^\mu, \nn\\
{\Aa}^{\seed}=&{\Aa}^{\seed}_{M} {\d}x^M=- \frac{\sqrt{3} ({{\tQ}}_0+x^\alpha \partial_\alpha {{\tQ}}{\0})} {r^2}dt + \ve \frac{\sqrt{3}{{\tQ}}_0}{r^2} (x^\alpha\partial_\alpha u_i{\0}) {\d}x^i,
\end{align}
where
\begin{equation}
f_0(r)=1- \frac{M_0}{r^4} +\frac{Q^2_0 +{{\tQ}}^2_0}{r^6},\qquad
f_1(r,x)=-\frac{x^\alpha \partial_\alpha M{\0}}{r^4}+ \frac{2Q_0 x^\alpha \partial_\alpha Q{\0}+ 2{{\tQ}}_0 x^\alpha \partial_\alpha {{\tQ}}{\0}}{r^6}.
\end{equation}
As explained below (\ref{promotion}), in order to satisfy the equations of motion \eqref{Einstein}-\eqref{MaxwellA},
suitable corrections must be added on top of (\ref{seed}).
Up to the first-order in the derivative expansion,
\begin{equation} \label{correction1}
\begin{split}
ds^2_{\corr}&\equiv g_{MN}^{\corr}{\d}x^M {\d}x^N =\frac{k(r)}{r^2}{\d}t^2+ 2 h(r) {\d}t{\d}r+ \frac{2j_i{\br}}{r^2} dt {\d}x^i
+r^2\Big[\alpha_{ij}{\br} -\frac{2}{3}\delta_{ij} h(r)\Big]{\d}x^i{\d}x^j,\\
{\Av}^{\corr}&={\Av}^{\corr}_M{\d}x^M= {\av}_t(r)dt+{\av}_i{\br}{\d}x^i,\qquad~~
{\Aa}^{\corr}={\Aa}^{\corr}_M{\d}x^M={\aa}_t(r)dt+{\aa}_i{\br}{\d}x^i,
\end{split}
\end{equation}
where $\alpha_{ij}{\br}$ is a traceless symmetric tensor of rank two under $SO(3)$ rotational symmetry between the spatial directions $x^i$.
In the parameterizing of the corrections (\ref{correction1}), we choose the following gauge convention
\begin{align} \label{gauge-fluid}
g^{\corr}_{rr}=0,\qquad g^{\corr}_{r\mu}\propto u_\mu,\qquad \textrm{Tr}\left[(g^{\textrm{eq}})^{-1}g^{\corr}\right]=0,\qquad
{\Av}^{\corr}_r={\Aa}^{\corr}_r=0.
\end{align}

Plugging the total bulk metric and gauge fields
\begin{align}
g_{MN}=g_{MN}^{\seed}+g_{MN}^{\corr}, \qquad
{\Av}_M={\Av}^{\seed}_M +{\Av}^{\corr}_M, \qquad
{\Aa}_M={\Aa}^{\seed}_M +{\Aa}^{\corr}_M,
\end{align}
into \eqref{Einstein}-\eqref{MaxwellA} results in a system of ordinary differential equations for those corrections in (\ref{correction1}). We need to specify suitable boundary conditions in order to fully determine the corrections. The first type of boundary condition is the requirement of asymptotic AdS, which fixes the large $r$ behavior for the corrections,
\begin{equation} \label{asymp requirement}
\begin{split}
k(r)&< \o(r^4),{\qquad} h(r)<\o(r^0), {\qquad} j_i{\br}<\o(r^4), {\qquad} \alpha_{ij}{\br} <\o(r^0),\\
{\av}_t(r)&<\o(r^0),{\qquad} {\aa}_t(r)<\o(r^0), {\qquad} {\av}_i{\br}<\o(r^0),{\qquad} {\aa}_i{\br}<\o(r^0).
\end{split}
\end{equation}
The second type of boundary condition is the regularity requirement for all the corrections in (\ref{correction1}),
\begin{equation} \label{regularity}
h{\br},\, k{\br},\, j_i{\br},\, \alpha_{ij}{\br},\,
{\av}_t{\br},\,  {\av}_i{\br},\,  {\aa}_t{\br},\,  {\aa}_i{\br}~\textrm{are regular over}~r\in[\rh ,+\infty),
\end{equation}
which turns out to be effective at the event horizon $r=\rh $. The remaining ambiguity of determining the corrections of (\ref{correction1}) will be fixed by the frame convention. We will work in Landau-Lifshitz frame so that
\begin{equation} \label{Landau frame}
u_\mu {T}^{\mu\nu}=- \E u^\nu,\qquad
u_\mu {J}^\mu =-{\rn},\qquad
u_\mu {J}^\mu_5 =-{\rn}_5,
\end{equation}
where $\E$, ${\rn}$, ${\rn}_5$ are the energy density, vector and axial charge densities of the fluid, respectively,
\begin{equation} \label{energy-charge}
\E= 3M(x),\qquad {\rn}=2\sqrt{3}Q(x), \qquad {\rn}_5=2\sqrt{3}{{\tQ}}(x).
\end{equation}
Note the identification made in (\ref{energy-charge}) is the promotion from their in-equilibrium counterparts (\ref{eq energy-charge}). Up to the first-order in the derivative expansion, the Landau-Lifshitz frame conditions (\ref{Landau frame}) turn into constraints on the form of boundary stress-energy tensor and currents,
\begin{align}
{T}_{tt} &= 3M_0+3x^\alpha \partial_\alpha M{\0},\label{Landau Ttt}\qquad\quad
{T}_{ti}  = {T}_{it}=-4M_0 (x^\alpha\partial_\alpha u_i{\0}),\\ 
{J}^t &= 2\sqrt{3}\left[Q_0+x^\alpha\partial_\alpha Q{\0}\right],\label{Landau Jt}\qquad
{J}^t_5 = 2\sqrt{3}\left[{{\tQ}}_0+x^\alpha\partial_\alpha {{\tQ}}{\0}\right]. 
\end{align}

For the sake of later presentation, we rewrite the expressions \eqref{Tmunu}-\eqref{J5mu} in terms of those corrections in (\ref{correction1}),
\begin{align}
{T}_{tt}&=3M_0+3x^\mu \partial_\mu M{\0}-2r^3 (\partial_k u_k{\0})+6r^4h(r)+2r^5 \partial_r h(r) +3k(r),\label{Ttt}\\
{T}_{ti}&={T}_{it}=-r^3\partial_t u_i{\0}+4j_i{\br}-r\partial_r j_i{\br} -4M_0 (x^\alpha\partial_\alpha u_i{\0}), \label{Tti}\\
{T}_{ij}&=\big[ M_0+x^\alpha \partial_\alpha M{\0}+k(r)-r\partial_r k{\br}-6r^4 h(r)+\frac{4}{3}r^3 (\partial_k u_k{\0})\big] \,\delta_{ij}\nn \\
&\quad -r^3 \big[\partial_i u_j{\0}+\partial_j u_i{\0} -\frac{2}{3} \delta_{ij}(\partial_k u_k{\0}) \big] -r^5\partial_r \alpha_{ij}{\br}, \label{Tij}
\end{align}
as well as
\begin{align} \label{Jti}
{J}^t&=2\sqrt{3}(Q_0+x^\alpha \partial_\alpha Q{\0})+r^3 \partial_r {\av}_t , \qquad
{J}^i=2\sqrt{3}Q_0 (x^\alpha\partial_\alpha u_i{\0})-r^3\partial_r {\av}_i{\br} -r {F}_{ti}^{\rme}{\0},\\
\label{J5ti}
{J}^t_5&=2\sqrt{3}({{\tQ}}_0+x^\alpha \partial_\alpha {{\tQ}}{\0})+r^3 \partial_r {\aa}_t ,\qquad
{J}^i_5=2\sqrt{3}{{\tQ}}_0 (x^\alpha\partial_\alpha u_i{\0})-r^3\partial_r {\aa}_i{\br}.
\end{align}
The limit of $r\to \infty$ is assumed implicitly in expressions above \eqref{Ttt}-\eqref{J5ti},
and we have also ignored the terms that will be explicitly vanishing as $r\to \infty$.

While the bulk corrections will be constructed around $x^\mu=0$, they do contain enough information to write down the total bulk metric and gauge fields about any point, valid up to the first-order in the derivative expansion. Instead of following this approach of \cite{Bhattacharyya:2008jc}, we will compute the boundary stress-energy tensor and currents using thus-constructed solutions, via the formulas \eqref{Ttt}-\eqref{J5ti}. Eventually, we will lift up thus-obtained constitutive relations into a covariant form.

\subsection{First-Order Charged Fluid: CESE and Other Conductivities} \label{constitutive}

Following Section \ref{setup fluid/gravity}, it is straightforward to solve the bulk equations \eqref{Einstein}-\eqref{MaxwellA} and obtain the corrections in \eqref{correction1}. In what follows, we summarize the final results and leave all the calculation details in Appendix \ref{appF-solving}. We present by grouping them into different sectors under $SO(3)$ symmetry of the boundary spatial directions.

In the scalar sector,
\begin{equation} \label{scalar sol}
k{\br}=\frac{2}{3}r^3 (\partial_k u_k{\0}),\qquad
h{\br}=0,\qquad {\av}_t{\br}=0,\qquad {\aa}_t{\br}=0.
\end{equation}

In the vector sector, thanks to going beyond the probe limit, dynamical equations for $j_i,\av_i,\aa_i$ are coupled, see \eqref{eom vi}-\eqref{eom ji}. It is exactly this coupling that makes CESE non-vanish, as opposed to the probe limit \cite{Bu:2016oba, Bu:2016vum}.
Since the final solutions in the vector sector are very lengthy, we record their near boundary expansions only,
\begin{align}
{\av}_i{\br}\xrightarrow[]{r\to\infty}&\,- \frac{3\sqrt{3}Q_0} {r^2} \Big(\frac{\rh ^3}{4M_0}+\frac{1}{4\rh }\Big) \partial_tu_i{\0}- \frac{\sqrt{3}}{r^2} \frac{(2\rh ^6+Q_0^2+2{{\tQ}}_0^2)}{4M_0\rh ^3}\partial_iQ{\0}+\frac{\sqrt{3}}{r^2}\frac{Q_0{{\tQ}}_0}{4M_0\rh ^3} \partial_i{{\tQ}}{\0}    \nn \\
& + \frac{1}{r} {F}_{ti}^{\rme}{\0} - \frac{1}{r^2} \Big(\frac{3Q_0^2}{4M_0\rh }+\frac{1}{2}\rh \Big){F}_{ti}^{\rme}{\0}+\o\(r^{-3}\),  \label{large r vi} \\
 \label{large r ai}
{\aa}_i{\br}\xrightarrow[]{r\to\infty}&- \frac{3\sqrt{3}{{\tQ}}_0}{r^2} \Big(\frac{\rh ^3} {4M_0}+\frac{1}{4\rh }\Big) \partial_tu_i{\0}- \frac{\sqrt{3}}{r^2} \frac{(2\rh ^6+2Q_0^2+{{\tQ}}_0^2)}{4M_0\rh ^3}\partial_i{{\tQ}}{\0} +\frac{\sqrt{3}}{r^2}  \frac{Q_0{{\tQ}}_0}{4M_0\rh ^3} \partial_iQ{\0}  \nn\\ &
- \frac{1}{r^2} \frac{3Q_0{{\tQ}}_0}{4M_0\rh }{F}_{ti}^{\rme}{\0}+\o\(r^{-3}\), \\
 \label{large r ji}
j_i{\br}\xrightarrow[]{r\to\infty}& \, r^3\partial_tu_i{\0}+\o (r^{-1}).
\end{align}

The tensor sector is the most simple one
\begin{align} \label{large r alphaij}
\alpha_{ij}{\br}
= & 3 \left[\partial_iu_j{\0}+ \partial_ju_i{\0}-\frac{2}{3}\delta_{ij}(\partial_k u_k{\0}) \right] \int_\infty^r \frac{{\d}\x}{\x^5f_0(\x)} \int_{\rh }^{\x} y^2{\d}y,\nn \\
\xrightarrow[]{r\to\infty} & \left(\frac{1}{r}-\frac{\rh ^3}{4r^4}\right) \left[\partial_iu_j{\0}+ \partial_ju_i{\0}-\frac{2}{3}\delta_{ij}(\partial_k u_k{\0}) \right] +\o(r^{-5}).
\end{align}

Now it is direct to compute the boundary stress-energy tensor and currents by substituting \eqref{scalar sol}-\eqref{large r alphaij} into \eqref{Ttt}-\eqref{J5ti}. Once lifted up into covariant form, the boundary stress-energy tensor $T_{\mu\nu}$ is given by \eqref{stress-cov}\eqref{pimunu}
\begin{align} \label{covariant Tmunu}
{T}_{\mu\nu}= {\E} u_\mu u_\nu+\P P_{\mu\nu}-2\eta \sigma_{\mu\nu}- \zeta (\partial_\alpha u^\alpha)P_{\mu\nu} + \cdots ,
\end{align}
where the projection tensor $P_{\mu\nu}$ and shear tensor $\sigma_{\mu\nu}$ are defined as
\begin{align}
 P_{\mu\nu}=\eta_{\mu\nu}+u_\mu u_\nu,\qquad
\sigma_{\mu\nu}\equiv \frac{1}{2}P_\mu^\alpha P_\nu^\beta \left(\partial_\alpha u_\beta +\partial_\beta u_\alpha\right)-\frac{1}{3}P_{\mu\nu} (\partial_\alpha u^\alpha).
\end{align}
${\E}$ is the energy density and ${\P}$ is the pressure of the fluid, which satisfy ${\E}=3{\P}$ and
\begin{align} \label{energy-pressure field}
{\E}=3M(x),\qquad {\P}=M(x).
\end{align}
The shear viscosity $\eta$ and bulk viscosity $\zeta$ are 
\begin{equation}
\eta =\rh(x)^3, \qquad \qquad  \zeta=0.
\end{equation}
From \eqref{entropy density}, entropy density of the dual fluid is $s=4\pi \rh(x)^3$. So, as expected, our result for shear viscosity $\eta$ saturates the Kovtun-Son-Starinets (KSS) bound \cite{Policastro:2001yc,Kovtun:2004de,Son:2007vk}.


Plugging \eqref{scalar sol}-\eqref{large r ai} into \eqref{Jti}-\eqref{J5ti} generates boundary currents \eqref{Jt+J5t pre-covariant}-\eqref{J5i pre-covariant}, which are, however, parameterized in terms of bulk quantities $Q(x),\tilde{Q}(x),r_h(x),M(x)$. Physically, we have to re-parameterize \eqref{Jt+J5t pre-covariant}-\eqref{J5i pre-covariant} via boundary fluid-dynamical variables. Moreover, the chemical potentials will be preferred in expressing the diffusive terms of $J^\mu,J_5^\mu$. In what follows we outline the strategy of implementing this transformation but defer technical details to Appendix \ref{appF-relations}.

The vector and axial charge densities are promoted versions of \eqref{eq energy-charge}:
\begin{align} \label{density field}
 {\rn}(x) &= 2\sqrt{3}Q(x) ,  \qquad     {\rn}_5(x)= 2\sqrt{3}\tQ(x).
\end{align}
In the fluid/gravity correspondence, chemical potentials are defined as
\begin{equation} \label{define chemical potential}
\begin{split}
\mu(x) &\equiv {\Av}_t |_{r=\infty}-{\Av}_t|_{r=\rh (x)}=\frac{\sqrt{3}Q(x)}{\rh(x) ^2},\\
\mu_5(x) &\equiv {\Aa}_t |_{r=\infty}-{\Aa}_t|_{r=\rh (x)}=\frac{\sqrt{3}{{\tQ}}(x)}{\rh(x) ^2}.
\end{split}
\end{equation}
From \eqref{Hawking temperature}, the temperature field of the dual fluid is
\begin{align} \label{temperature field}
T= \frac{\rh(x) }{\pi} \left(1-\frac{Q(x)^2+ {\tQ}(x)^2}{2\rh(x)^6}\right).
\end{align}
Using \eqref{energy-pressure field}, \eqref{density field}-\eqref{temperature field}, one can check that the following relation still holds
\begin{align}
{\E} +{\P} =T s +\mu  {\rn}  +\mu_5 {\rn}_5.
\end{align}

The relations \eqref{define chemical potential}-\eqref{temperature field} are useful in expressing $\partial_iQ$, $\partial_i\tilde{Q}$ and $\partial_i \rh$ in terms of $\partial_iT/T$, $\partial_i\mu$ and $\partial_i \mu_5$, see \eqref{temperature derivative}-\eqref{mu5 derivative} in Appendix \ref{appF-relations}.
Eventually, \eqref{Jt+J5t pre-covariant}-\eqref{J5i pre-covariant} could be recast into covariant forms,
which cover non-anomalous part of \eqref{covariantJm}\eqref{covariantJm5} with all transport coefficients expressed in terms of bulk parameters
\begin{align}
\svq&=\rh + \frac{3Q^2}{2M\rh}-\frac{9Q^2}{2M}\left(\frac{\rh^3}{2M}+ \frac{1}{2\rh} \right),\qquad\qquad  \stq=\svq\big|_{Q\leftrightarrow \tilde Q}, \label{sigmae-bulk}\\
\saq&=\frac{3Q\tilde{Q}}{2M\rh}- \frac{9Q\tilde{Q}}{2M} \left( \frac{\rh^3}{2M} +\frac{1}{2\rh} \right), \qquad \qquad \quad \staq=\saq. \label{sigma5e-bulk}
\end{align}
Note independent transport coefficients are $\svq$ and $\saq$, which are ``quantum critical'' or ``incoherent'' conductivities of hydrodynamics as in \cite{Hartnoll:2007ih,Hartnoll:2014lpa,Davison:2015taa}. This is partly due to the time-reversal symmetry of our holographic model.

Physically, bulk quantities in \eqref{sigmae-bulk}\eqref{sigma5e-bulk} should be eliminated in favor of thermodynamic variables of the boundary fluid. There are several ways of presenting the results.
When discussing single charge limit or probe limit, we find it more transparent to split the conductivities \eqref{sigmae-bulk}\eqref{sigma5e-bulk} into two parts, represented by $\sigma_Q$ and $\sigma_0$, as flashed in \eqref{sigma=sigmaQ+sigma0}\eqref{sigmaQ0}. On the other hand, in comparison with relevant works \cite{Huang:2013iia,Jiang:2014ura,Pu:2014cwa,Pu:2014fva}, we express $\svq$ and $\saq$ as functions of chemical potentials $\mu,\mu_5$ and fluid temperature $T$, as summarized in \eqref{sigma_mumu5T}. We also recast the hydrodynamic constitutive relations \eqref{covariantJm}\eqref{covariantJm5} into linear response form, see \eqref{Jmu_linear} \eqref{Jmu5_linear}. Then, we give a clarification for the differences between physical conductivities directly read off from Kubo formulas and intrinsic ones parameterizing hydrodynamic constitutive relations.

\section{Holographic AC Conductivities from Linear Response Analysis } \label{LinearAnalysis}

As a complementary study of Section \ref{FluidGravity}, in this section, we reveal some transport phenomena of the holographic model \eqref{bulkaction} through linear response analysis. We focus on the vector and axial currents generated by the external vector and axial electric fields and thermal gradient. We assume these external sources are weak in amplitudes and oscillate in time only.

\subsection{Black Brane Fluctuations and Conductivity Matrix} \label{fluc-conduc}

To study linear response transports within the holographic framework, we follow standard procedure and perturb the homogeneous black brane \eqref{homogeneous}
\begin{equation} \label{brane-fluctuation}
g_{MN}= g_{MN}^{\eq}+\delta g_{MN}, \qquad
{\Av}_M= {\Av}_M^{\eq}+ \delta {\Av}_M, \qquad
{\Aa}_M= {\Aa}_M^{\eq}+\delta {\Aa}_M.
\end{equation}
Diffeomorphism and $U(1)$ gauge invariance in the bulk theory allow choosing a particular gauge. Different from \eqref{gauge-fluid}, throughout this section, we will work under radial gauge convention,
\begin{equation}
\delta g_{rA}=0,\qquad  \qquad
\delta {\Av}_r=0,\qquad  \qquad \delta {\Aa}_r=0.
\end{equation}
Black brane fluctuations \eqref{brane-fluctuation} could be classified into decoupled sectors \cite{Kovtun:2005ev} according to their transformation properties under the remaining symmetry group $SO(3)$. For the purpose of computing electrical and thermal conductivities, we consider the helicity one sector only
\begin{equation} \label{helicity-one}
\delta (ds^2)=\epsilon\, 2\delta g_{ti}(r,t) dt {\d}x^i,~~~~~~\delta {\Av}= \epsilon\, {\delta{\Av}_i}(r,t) {\d}x^i,~~~~~~\delta {\Aa}= \epsilon\, {\delta{\Aa}_i}(r,t) {\d}x^i,
\end{equation}
where $\epsilon$ is a formal parameter marking the linearization. The calculations below will be accurate up to $\mathcal{O}(\epsilon^1)$, as required for linear response study.

Fluctuations \eqref{helicity-one} satisfy a system of partial differential equations, whose derivation is presented in Appendix \ref{appL}, see \eqref{Eri}-\eqref{Eri-Eti}. While there is no explicit interaction term between ${\Av}$ and ${\Aa}$ in the bulk action (\ref{bulkaction}), fluctuations ${\delta{\Av}_i}$ and ${\delta{\Aa}_i}$ do interact via $\delta g_{ti}$. As in Section \ref{FluidGravity}, this is exactly our point that the CESE can be realized by going beyond probe limit in a simple holographic model.

We proceed by deriving the compact forms of stress-energy tensor and currents of the boundary theory. To this end, we solve \eqref{Eri}-\eqref{EAi} near conformal boundary $r=\infty$,
\begin{align} \label{asymptotic delg}
\delta g_{ti}(r,t) &\xrightarrow[]{r\to \infty} r^2{\dhti}(t)+ \frac{\delta h_{ti}^{(4)}(t)} {r^2}+\o\(r^{-3}\),\\
 \label{asymptotic delV}
{\delta{\Av}_i}(r,t) &\xrightarrow[]{r\to\infty} {\davi}(t)+\frac{\partial_t{\davi}(t)}{r}+ \frac{\delta{\av}_i^{(2)}(t)}{r^2} -\frac{\log r}{2r^2} \partial_t^2 {\davi}(t)+\o\Big(\frac{\log r}{r^3}\Big),\\
 \label{asymptotic delA}
{\delta{\Aa}_i}(r,t) & \xrightarrow[]{r\to\infty} {\daai}(t)+\frac{\partial_t{\daai}(t)}{r}+ \frac{\delta{\aa}_i^{(2)}(t)}{r^2} -\frac{\log r}{2r^2} \partial_t^2 {\daai}(t)+\o\Big(\frac{\log r}{r^3}\Big),
\end{align}
where the non-normalizable modes ${\dhti}(t)$, ${\davi}(t)$, ${\daai}(t)$ correspond to external sources, see \eqref{thermal-metric}-\eqref{vector/axial E} for the precise identification. The constraint equation (\ref{Eri-Eti}) yields
\begin{equation} \label{constraint delg0V0A0}
4\delta h_{ti}^{(4)}=2\sqrt{3} ({{\tQ}}{\daai}+Q{\davi} ).
\end{equation}
The rest normalizable modes $\delta{\av}_i^{(2)}$, $\delta{\aa}_i^{(2)}$ have to be determined via fully solving the bulk equations \eqref{Eri}-\eqref{EAi}. Our strategy of solving them goes in two steps: (1) factorize out the time-dependence by basis decomposition \eqref{basis decomposition}; (2) solve ordinary differential equations satisfied by decomposition coefficients, see \eqref{eom S1}\eqref{eom S2}. As a result,
\begin{align}
\delta a_i^{(2)}= \frac{\mu s_1+ \mu_5 s_2}{ \mu^2+ \mu_5^2} \left(\mu \delta a_i^{(0)}  + \mu_5 \delta \tilde{a}_i^{(0)} \right), \qquad
\delta \tilde{a}_i^{(2)}= \frac{\mu_5 s_1- \mu s_2}{ \mu^2+ \mu_5^2} \left(\mu \delta a_i^{(0)}  + \mu_5 \delta \tilde{a}_i^{(0)} \right),
\end{align}
where $s_1,s_2$ encode pre-asymptotic behaviors of the decomposition coefficients, see \eqref{asymp expansion S1S2}.

With the near boundary expansions \eqref{asymptotic delg}-\eqref{asymptotic delA},
the dual stress-energy tensor and currents \eqref{Tmunu}-\eqref{J5mu} become,
\begin{align} \label{Tmunu linear}
T_{tt}&=3M, \qquad T_{ij}=M\delta_{ij}, \qquad
T_{ti}=M{\dhti}+2\sqrt{3}  ({{\tQ}}{\daai}+Q{\davi}  ),\\
J^t &=2\sqrt{3}Q,\qquad \,
J^i=2 \delta{\av}_i^{(2)}-\frac{1}{2} \partial_t^2 {\davi} - 2\sqrt{3} Q {\dhti}, \label{Jmu linear}\\
J^t_5 &=2\sqrt{3}{{\tQ}},\qquad
J^i_5=2\delta{\aa}_i^{(2)}-\frac{1}{2} \partial_t^2 {\daai} -2\sqrt{3} {{\tQ}} {\dhti}, \label{J5mu linear}
\end{align}
where the constraint relation (\ref{constraint delg0V0A0}) has been used to simplify $T_{ti}$. The heat current $\Q^i$ is defined as
\begin{equation} \label{heat current}
\Q^i=T^{ti}-\mu J^i-\mu_5 J_5^i,
\end{equation}
where $T^{ti}$ could be computed from $T_{ti}$ by a weakly curved boundary metric $\eta_{\mu\nu}+{\dhti}$. Thus, the currents \eqref{Jmu linear}-\eqref{heat current}, as the linear response to external sources $E_i,E_i^5,\nabla_i T$, can be summarized in a compact matrix form
\begin{equation} \label{conductivity matrix}
\left(
\begin{matrix}
J^i\\
J^i_5\\
\Q^i
\end{matrix}
\right)=
\left(
\begin{matrix}
\sv\bo&\st_5\bo&\alpha\bo T\\
\sa\bo&\st\bo&\alpha_5\bo T\\
\bar{\alpha}\bo{T} & \bar{\alpha}_5\bo{T}& \bar{\kappa}\bo{T}
\end{matrix}
\right)
\left(
\begin{matrix}
E_i\\
E_i^5\\
-\frac{\nabla_i T}{T}
\end{matrix}
\right),
\end{equation}
where \eqref{thermal-metric}-\eqref{vector/axial E} have been used.

The electrical conductivities are
\begin{equation}
\begin{split} \label{AC Conductivities}
\sv\bo &=-\Big[\underline{\pi\delta(\omega)}+\frac{i}{\omega}\Big]  \frac{2 (\mv^2s_1+ \ma^2s_2 )}{ (\mv^2+\ma^2 )}- \frac{1}{2}i\omega,\qquad \qquad
\st\bo=\sv\bo\big |_{\mv\leftrightarrow {\ma}}, \\
\sa\bo&=\st_5\bo=-\Big[\underline{\pi\delta(\omega)}+\frac{i}{\omega}\Big]   \frac{2\mv\ma(s_1-s_2)}{ (\mv^2+ \ma^2 )}.
\end{split}
\end{equation}
The thermoelectric conductivities are
\begin{equation}
\begin{split}  \label{ACalpha}
\alpha\bo{T} &=\bar{\alpha}\bo{T}=\Big[\underline{\pi\delta(\omega)}+\frac{i}{\omega}\Big]     {\rn}
- ( \mu \sigma + \mu_5 \st_5 ),\\
\alpha_5\bo{T}&=\bar{\alpha}_5\bo{T}  =\Big[\underline{\pi\delta(\omega)}+\frac{i}{\omega}\Big]    {{\rn}_5}  -
( \mu \sa\bo +  \mu_5 \st\bo).
\end{split}
\end{equation}
Finally, the thermal conductivity is
\begin{align}
\bar{\kappa}\bo{T} &=\Big[\underline{\pi\delta(\omega)}+\frac{i}{\omega}\Big]   \(\E+{\P}-2{\rn}\mv -2{\rn}_5\ma\) +
\(\mu^2\sv\bo +\mu_5^2 \st\bo+ 2 \mu \mu_5 \sa\bo\).\label{ACkappa}
\end{align}
As seen from \eqref{AC Conductivities}\eqref{ACalpha}\eqref{ACkappa}, only $\sigma$ and $\sa$ are independent: $\tilde\sigma$ could be extracted from $\sigma$ via the exchange $\mu\leftrightarrow\mu_5$; all remaining conductivities are determined by their combinations.
$\sv\bo$ and $\st\bo$ are the Ohmic electrical conductivities for vector current and axial current, and $\sa$ corresponds to the chiral electric separation effect while $\st_5$ is its vector analogue. The CESE conductivity $\sa$ is invariant under the exchange of $\mu$ and $\mu_5$.

In \eqref{ACalpha}, $\alpha$ and $\alpha_5$ are the thermoelectric conductivities of generating vector and axial currents, respectively. Thanks to time reversal symmetry, Onsager reciprocal relations $\bar{\alpha}=\alpha$ and $\bar{\alpha}_5=\alpha_5$ do hold, and the conductivity matrix in \eqref{conductivity matrix} is symmetric.
In \eqref{ACkappa}, $\bar\kappa$ is the heat conductivity. In the numerator of $\bar\kappa$, we have made the replacement $\E \to \E+\P=4M$ as done in \cite{Hartnoll:2009sz,Herzog:2009xv, Hartnoll:2007ip}.
This added $\P$ is actually a ``contact term'' \cite{Policastro:2002tn} due to translation invariance. Then, it is consistent with \eqref{heat-conduc-hydro} obtained by fluid/gravity calculations.

Note the appearance of $i/\omega$ in the conductivities \eqref{AC Conductivities}\eqref{ACalpha}\eqref{ACkappa}. From the Kramers-Kronig relation \cite{Hartnoll:2008vx,Hartnoll:2008kx}, this means there must be a delta function $\delta{(\omega)}$ in real parts of all the conductivities. However, in holographic models with spatial translational invariance, it is not easy to track the delta-peak directly. For consistency, we just added this $\delta{(\omega)}$ as underlined terms above \cite{Hartnoll:2008vx,Hartnoll:2008kx}.

In the low-frequency limit where $\omega/T \ll 1$, we obtained analytical expressions for $\sigma,\sa$, see \eqref{sigma DC1}\eqref{sigma5 DC1}. For the purpose of comparing with \eqref{sigmaVo}\eqref{sigmaAo}, we eliminate bulk parameters in \eqref{sigma DC1}\eqref{sigma5 DC1} in favor of thermodynamic quantities of the boundary theory. Eventually, \eqref{sigma DC1}\eqref{sigma5 DC1} turn into
\begin{align} \label{sigma DC}
\sv\bo&= \Big[\underline{\pi\delta(\omega)}+\frac{i}{\omega}\Big]  \frac{{\rn}^2}{\E+\P}+{\svq} +\cdots,\qquad 
\sa\bo = \Big[\underline{\pi\delta(\omega)}+\frac{i}{\omega}\Big] \frac{{\rn}^2_5}{\E+\P}+{\saq}  + \cdots,
\end{align}
where $\cdots$ denote higher powers in $\omega$ corrections,
$\svq$ and $\saq$ are given in \eqref{sigma=sigmaQ+sigma0}.
Obviously, aside from the subtle piece $\delta(\omega)$,  the AC conductivities \eqref{sigma DC} are in perfect agreement with the result from linear response \eqref{sigmaVo}\eqref{sigmaAo}.
In the single charge limit, \eqref{sigma DC} also stands in line with the two-point correlators of \cite{Ge:2008ak}.
And in the low-frequency limit, the real part of them give the intrinsic conductivities
$\lim_{\omega\to 0} {\rm{Re}}[\sv] =\svq$ and $\lim_{\omega\to 0} {\rm{Re}}[\sa] =\saq$.
The $\omega$-dependence of $\sigma$ and $\sa$ will be the focus of the next subsection.

\subsection{AC Conductivities: Numerical Plots}

\begin{figure}[h]
\centering
\includegraphics[scale=0.48]{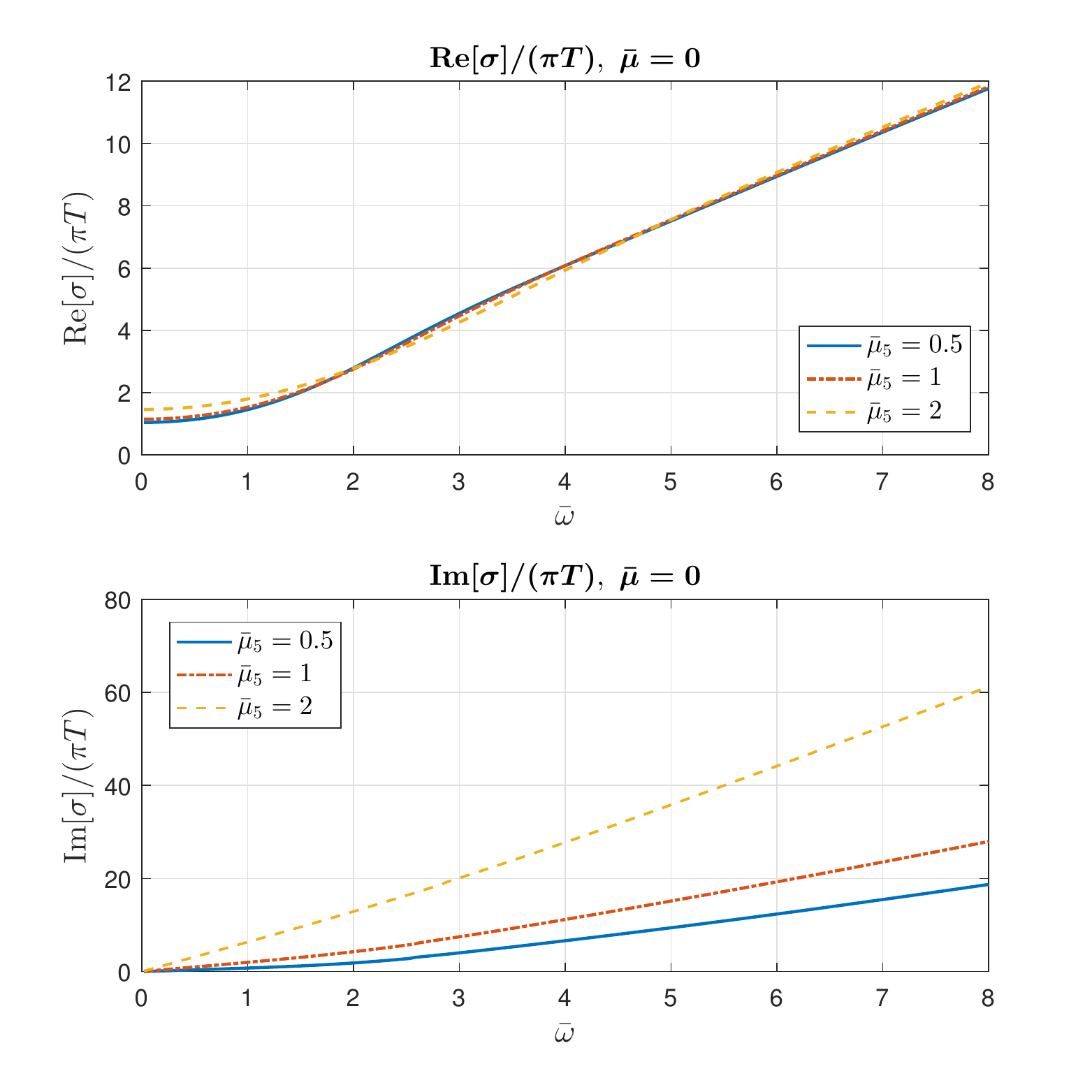}
\includegraphics[scale=0.48]{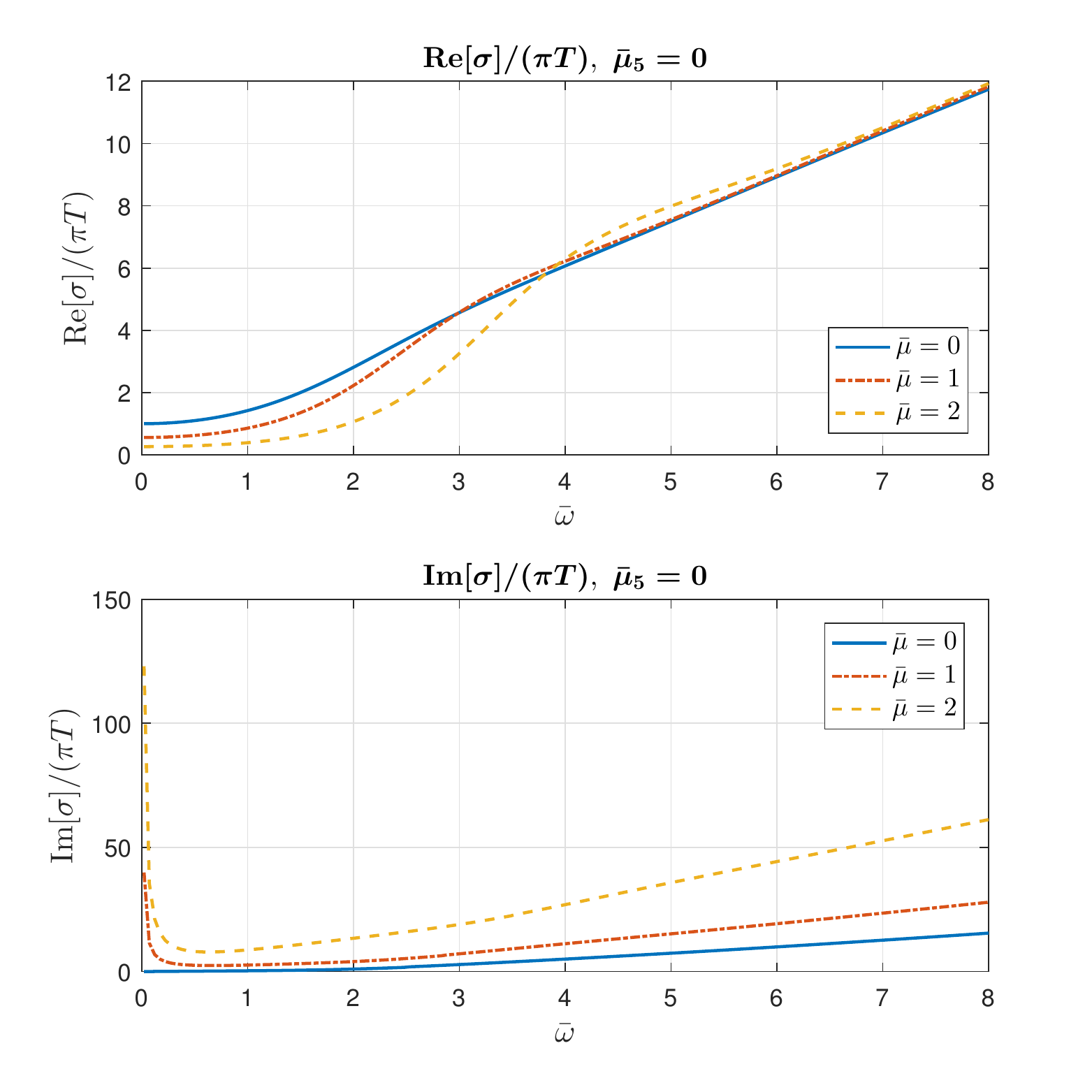}
\caption{Ohmic conductivity $\sigma/(\pi T)$ (Real and Imaginary parts) as a function of dimensionless frequency $\bar{\omega}=\omega/(\pi T)$ when either $\bar{\mu}=0$ (left) or $\bar{\mu}_5=0$ (right). Different curves correspond to different values of $\bar{\mu}_5=0.5, 1, 2$ (left) or $\bar{\mu}=0,1,2$ (right). From the right-bottom panel, there appears divergent behavior for imaginary part of $\sigma$: $\rm Im[\sigma]\sim 1/\bar\omega$. Via Kramers-Kronig relation, this implies there should be a delta peak in $\rm Re[\sigma]$ at $\bar\omega=0$, which we, however, could not track in numerical calculations.}
\label{sigma AC mumu5=0}
\end{figure}

\begin{figure}[h]
\centering
\includegraphics[scale=0.48]{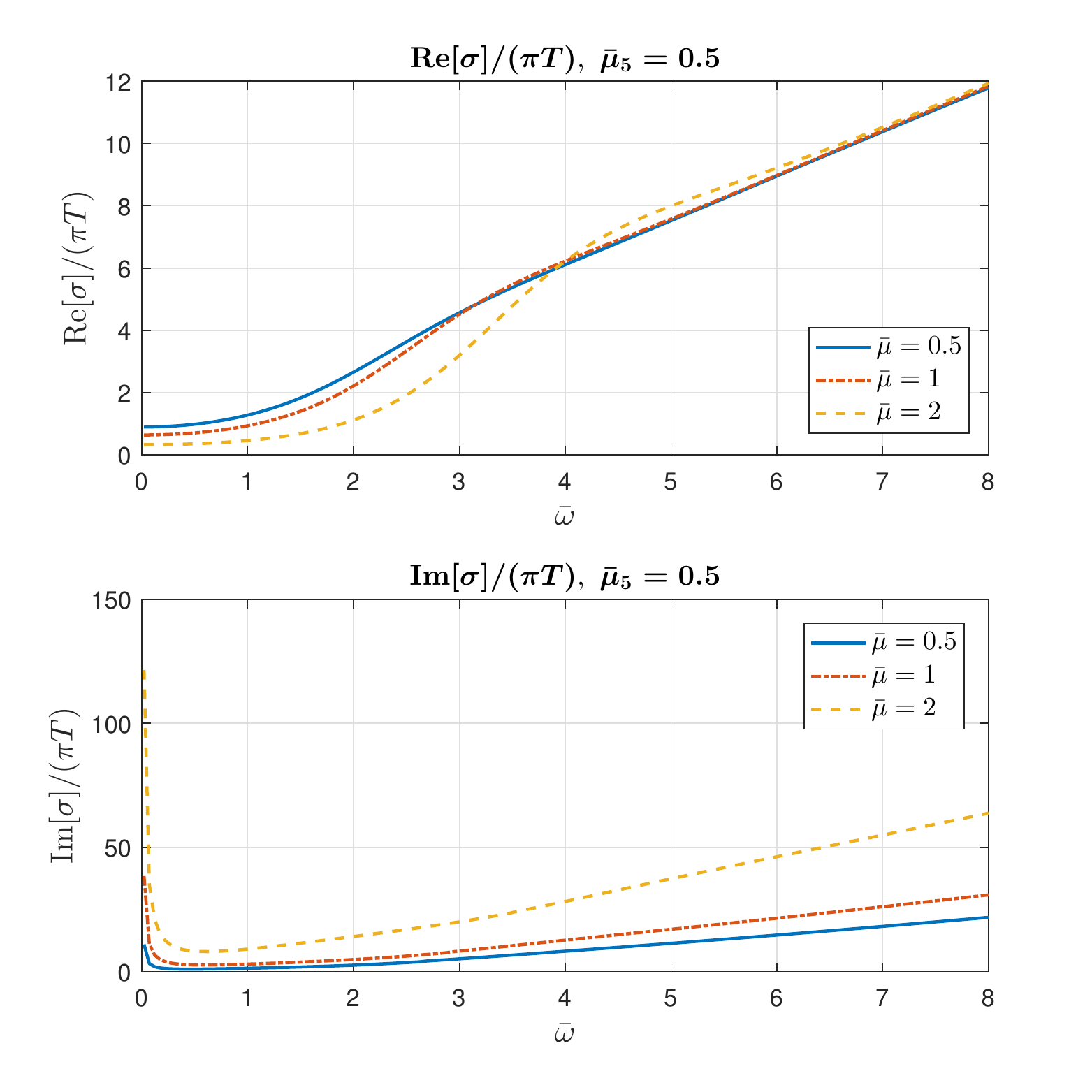}
\includegraphics[scale=0.48]{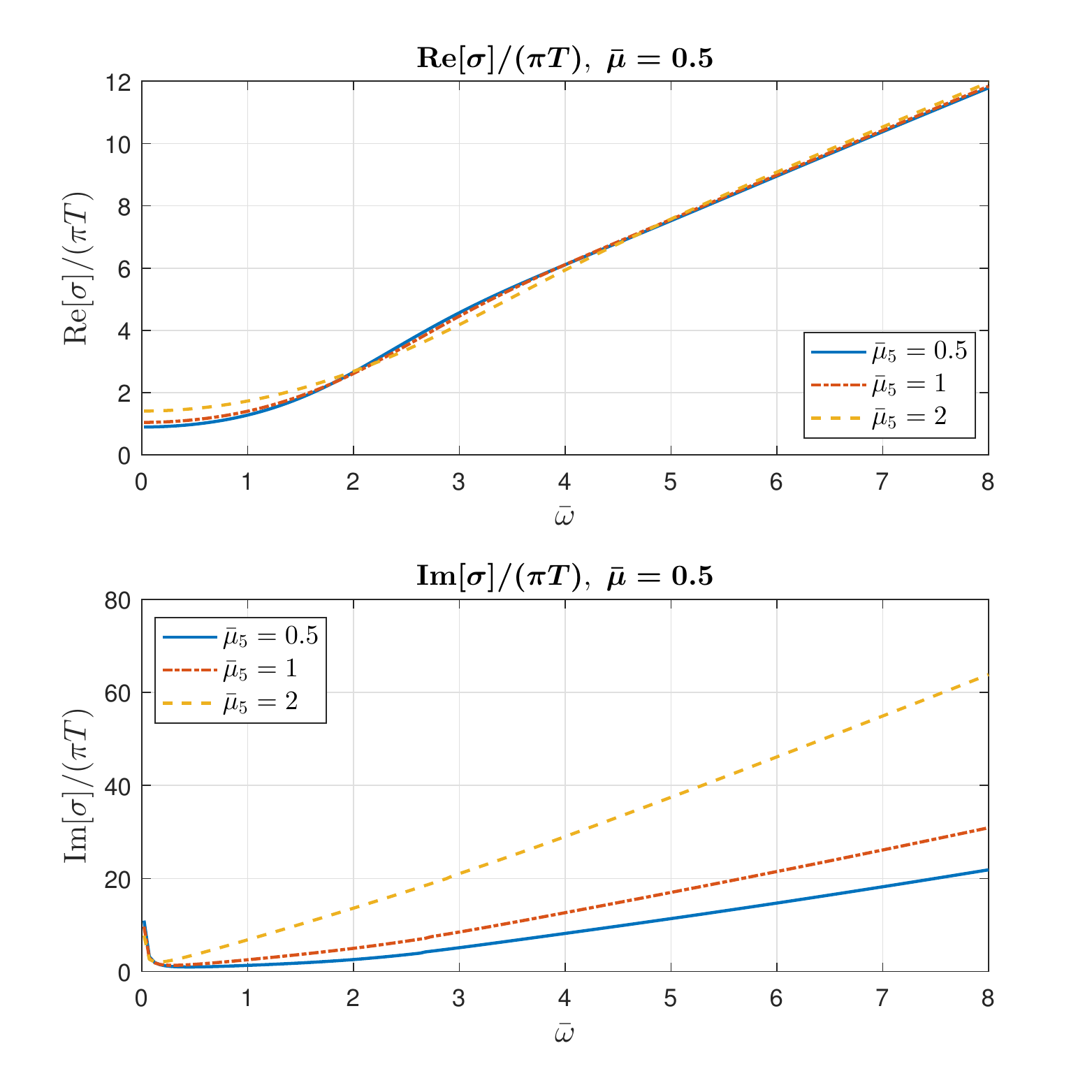}
\caption{Ohmic conductivity $\sigma/(\pi T)$ (Real and Imaginary parts) as a function of dimensionless frequency $\bar{\omega}=\omega/(\pi T)$ with fixed $\bar{\mu}_5=0.5$ (left) or $\bar{\mu}=0.5$ (right) . Different curves correspond to different choices of $\bar{\mu}=0.5, 1, 2$ (left) or $\bar{\mu}_5=0.5, 1, 2$ (right). As in Figure \ref{sigma AC mumu5=0}, the divergent behavior ($\sim 1/\bar\omega$) in $\rm Im[\sigma]$ implies the presence of a delta-peak at $\bar\omega=0$ in $\rm Re[\sigma]$, which we could not display numerically.}
\label{sigma AC mumu505}
\end{figure}

\begin{figure}[h]
\centering
\includegraphics[scale=0.48]{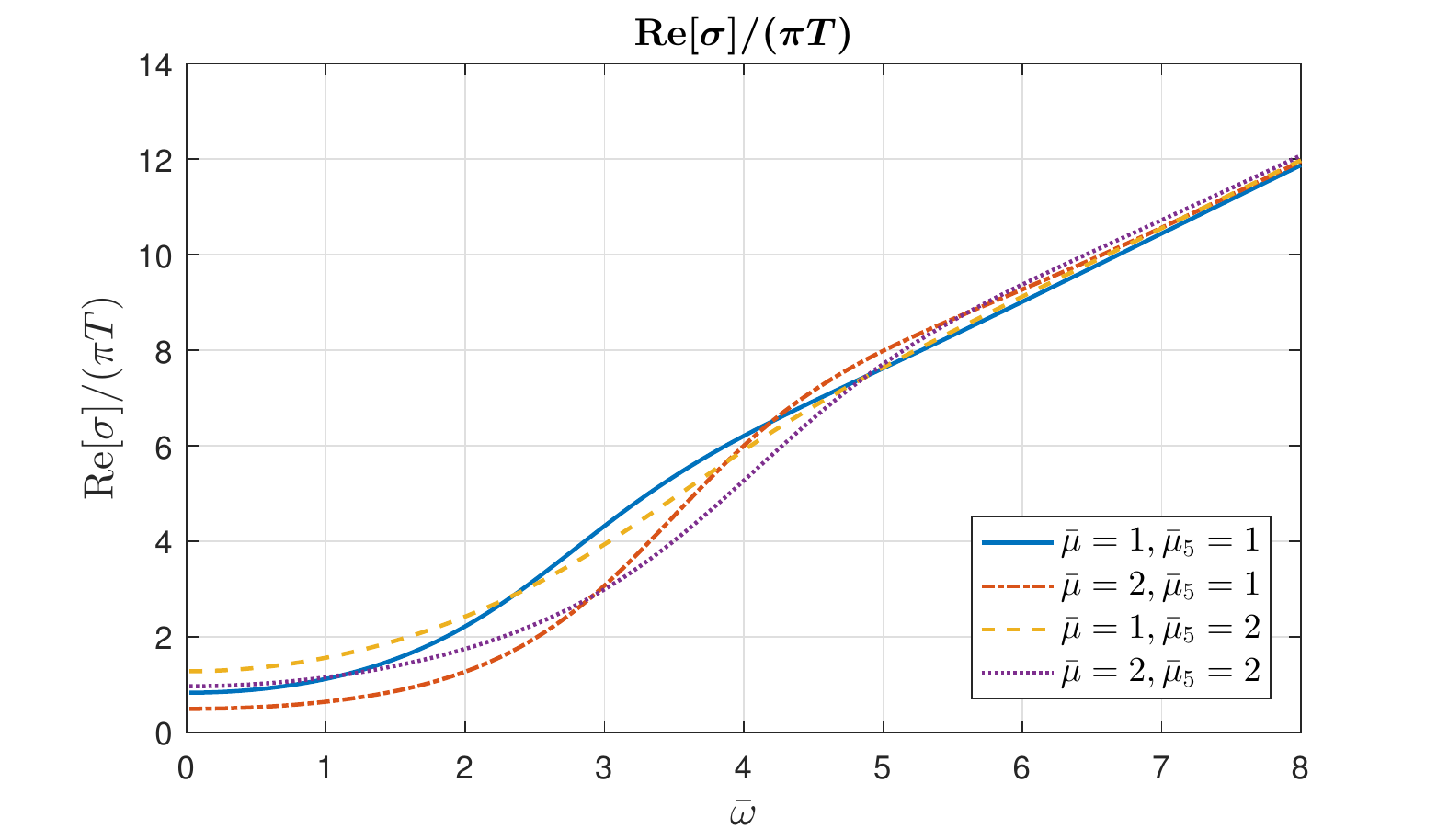}
\includegraphics[scale=0.48]{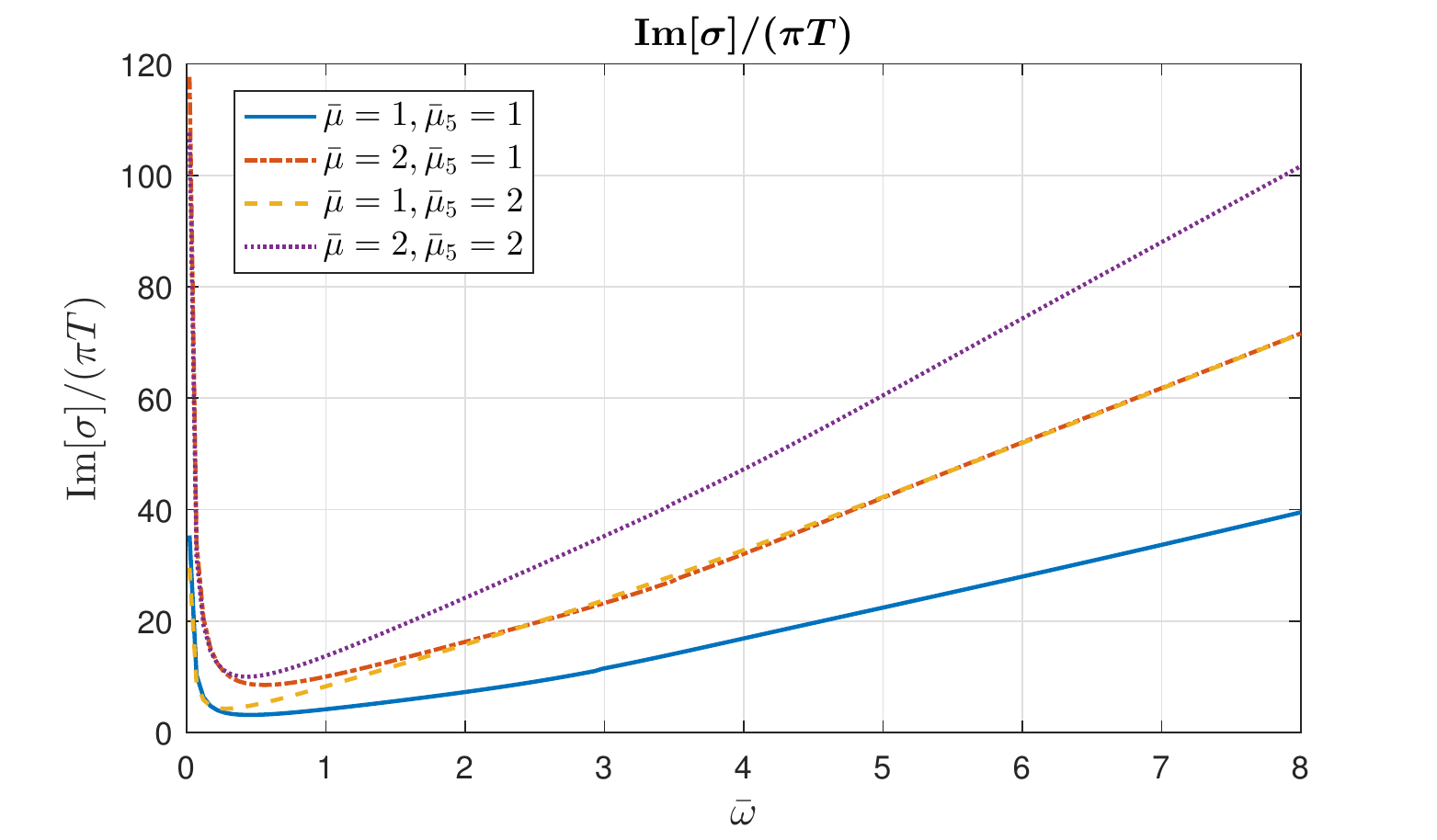}
\caption{Ohmic conductivity $\sigma/(\pi T)$ (Real and Imaginary parts) as a function of dimensionless frequency $\bar{\omega}=\omega/(\pi T)$ for larger values of $\bar{\mu}=1, 2$ and $\bar{\mu}_5=1,2$.  As in Figure \ref{sigma AC mumu5=0}, the divergent behavior ($\sim 1/\bar\omega$) in $\rm Im[\sigma]$ implies the presence of a delta-peak at $\bar\omega=0$ in $\rm Re[\sigma]$, which we could not display numerically.}
\label{sigma AC mumu5 larger}
\end{figure}

In this subsection, we present numerical results for the AC conductivities $\sigma$ and $\sigma_5$ while depositing more technical details in Appendix \ref{appL}.
Our results for frequency dependence of Ohmic conductivity $\sigma$ are plotted in Figures \ref{sigma AC mumu5=0}, \ref{sigma AC mumu505}, \ref{sigma AC mumu5 larger}.

Figure \ref{sigma AC mumu5=0} is about the plot of $\sigma/(\pi T)$ as a function of dimensionless frequency $\bar{\omega}$ when either $\bar{\mu}=0$ or $\bar{\mu}_5=0$.
As seen from left panels of Figure \ref{sigma AC mumu5=0}, when $\bar{\mu}=0$, i.e., neglecting the back-reaction effect from the vector field in the bulk, the low-frequency limit of $\sigma$ is always finite, which is in consistent with the analytical result (\ref{sigma DC}). For representative values of $\bar{\mu}_5$ chosen by us, increasing the axial chemical potential $\bar{\mu}_5$ results in reasonably profound modification for the infrared behaviors of $\sigma$. From the right-bottom panel of Figure \ref{sigma AC mumu5=0}, it is clear that, once $\bar{\mu}\neq 0$ (i.e. beyond probe limit),
there appears divergent behavior for imaginary part of $\sigma$: ${\rm{Im}}[\sigma]\sim {\bar{\omega}^{-1}}$, which is also in agreement with analytical result in (\ref{sigma DC}). Via Kramers-Kronig relation, this $1/\bar\omega$-behavior in $\rm Im[\sigma]$ means that ${\rm{Re}}[\sigma]\sim\delta({\bar{\omega}})$.
Moreover, the strength of back-reaction due to vector field in the bulk is also reflected by different curves in the right-bottom panel in Figure \ref{sigma AC mumu5=0}.

Figure \ref{sigma AC mumu505} is to further explore the effects of chemical potentials on $\sigma$. In the infrared regime of $\omega$, $\rm{Im}[\sigma]$ is dominated by the divergent behavior $\sim 1/\bar\omega$, which implies a delta-peak $\delta(\omega)$ for $\rm Re[\sigma]$. While we could not display this $\delta(\omega)$ in numerical plots, we find the finite piece in $\rm Re[\sigma]$ and particularly that $\lim_{\omega\to 0} {\rm{Re}}[\sigma] =\sigma_e$. Let us focus on the infrared regime (roughly with $0 <  \bar{\omega} \lesssim 2$) of $\rm{Re}[\sigma]$.
Our observation is that increasing $\bar{\mu}$ will diminish $\rm{Re}[\sigma]$ while increasing $\bar{\mu}_5$ will result in enhancement of $\rm{Re}[\sigma]$.
This is exactly consistent with the DC limit $\svq$ in \eqref{sigma_mumu5T}, which has  been plotted in the left panel of Figure \ref{plot sigmae+sigma5e}. The behavior for larger $\bar{\omega}$ is supposed to be controlled by UV conformal symmetry. To confirm this observation, in Figure \ref{sigma AC mumu5 larger} we plot Ohmic conductivity $\sigma$, as a function of $\bar{\omega}$, for larger chemical potentials $\bar{\mu}$ and $\bar{\mu}_5$. Roughly, when $\bar\omega\gtrsim 6$, $\rm{Re}[\sigma]$ becomes insensitive to change of chemical potentials.

\begin{figure}[h]
\centering
\includegraphics[scale=0.48]{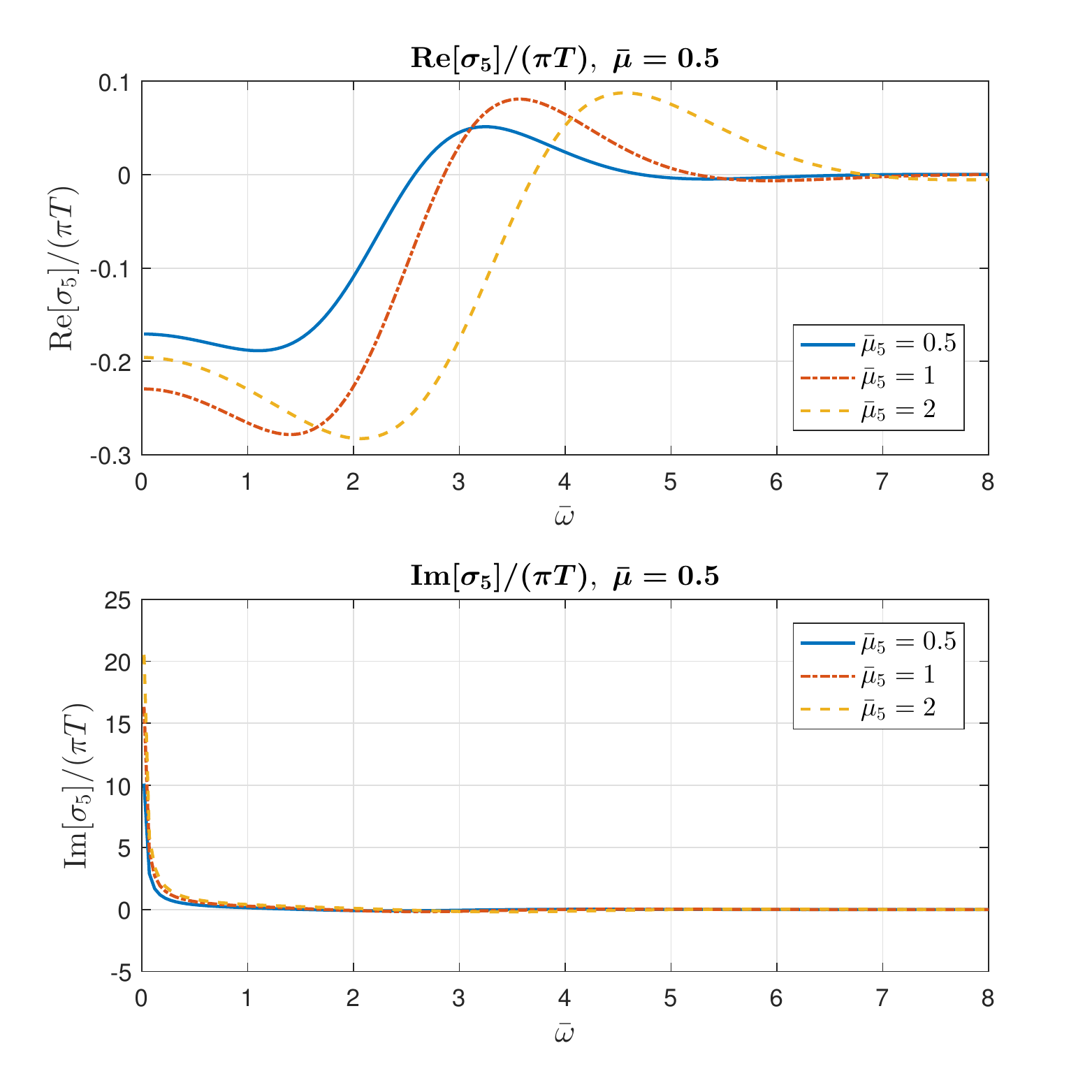}
\includegraphics[scale=0.48]{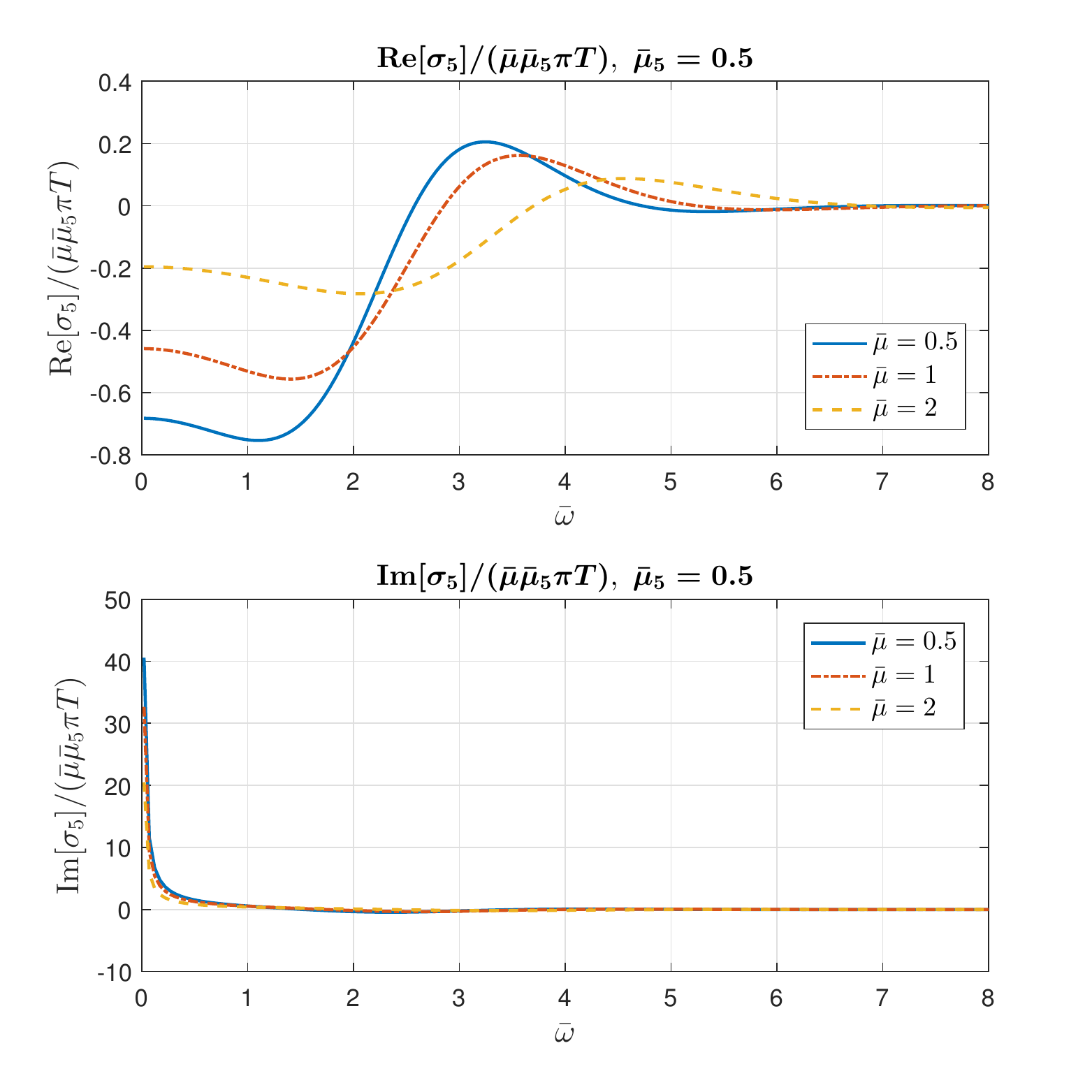}
\caption{Left: CESE conductivity $\sa/(\pi T)$ (Real and Imaginary parts) as a function of dimensionless frequency $\bar{\omega}=\omega/(\pi T)$. Right: The deviation factor $\sa/(\bar{\mu}\bar{\mu}_5 \pi T)$ (Real and Imaginary parts) as a function of $\bar{\omega}$. As in Figure \ref{sigma AC mumu5=0}, the divergent behavior ($\sim 1/\bar\omega$) in $\rm Im[\sigma_5]$ implies the presence of a delta-peak at $\bar\omega=0$ in $\rm Re[\sigma_5]$, which we could not display numerically.}
\label{sigma5 AC mumu5}
\end{figure}

Figure \ref{sigma5 AC mumu5} is to show the frequency-dependence of the CESE conductivity $\sigma_5$ and the generalized deviation factor $\chiA\equiv \sa/(\bar{\mu}\bar{\mu}_5 \pi T)$ extending \eqref{sigma_mumu5T} to AC case. Given that $\sigma_5$ is symmetric under the exchange of $\mu$ and $\mu_5$, it is legitimate to constrain to either $\mu \geq \mu_5$ or $\mu \leq \mu_5$ without loss of generality. Just like $\sigma$ displayed in Figures \ref{sigma AC mumu5=0}, \ref{sigma AC mumu505}, \ref{sigma AC mumu5 larger}, $\rm Im[\sigma_5]$ shows diverging behavior (i.e.$\sim 1/\omega$) near $\omega=0$, which, via Kramers-Kronig relation, indicates ${\rm{Re}}[\sigma]\sim\delta({\omega})$. Once away from $\omega=0$, the imaginary parts approach zero soon, while the real parts evolve in a more profound fashion as $\omega$ is increased.
Particularly, we observe a damped oscillating behavior for $\rm{Re}[\sigma_5]$, where the asymptotic regime is achieved around $\bar{\omega}=5.5$ for $\mu=0$. This is roughly in agreement with the numerical results of \cite{Pu:2014fva}, although we have the different mechanism of generating CESE.
Moreover, from Figure \ref{sigma5 AC mumu5} we observe that increasing $\mu$ or $\bar{\mu}_5$ would delay the achievement of the asymptotic regime.

\section{Conclusion and Discussions} \label{Conclusion}

In this work, we explored transport properties of strongly coupled matter, which is holographically described by $(4+1)$-dimensional Einstein gravity, coupled to $U(1)_V\times U(1)_A$ gauge fields, in the asymptotic AdS$_5$ black brane. Our main finding is the nonzero CESE conductivity when the gravitational back-reaction effect is taken into account, see 
\eqref{sigma_mumu5T} for the hydrodynamic limit and \eqref{AC Conductivities} for its extension to an AC conductivity.
We confirmed our results with two complementary studies --- fluid/gravity calculations versus linear response analysis. Within the former framework, we constructed the first-order constitutive relations for stress-energy tensor, vector and axial currents for the holographic matter in the long wavelength and low frequency limit. Following the linear response approach, we revealed the frequency-dependence of Ohmic, CESE and thermoelectric conductivities.

As the second task, we clarified the relations between the dissipative transport coefficients in the hydrodynamic constitutive relations \eqref{covariantJm}\eqref{covariantJm5} and those appearing in the conductivity matrix \eqref{conductivity matrix}. While the ``intrinsic'' conductivities $\svq,\saq$ etc. are widely used in the framework of fluid dynamics, the physical observable are indeed those appearing in the conductivity matrix \eqref{conductivity matrix}. Indeed, when the hydrodynamic description is reformulated into the linear response form, we find perfect agreement between these two different approaches, see \eqref{sigmaVo}\eqref{sigmaAo} \eqref{heat_linear} and \eqref{sigma DC}.

Since we have turned off the possible Chern-Simons terms in our holographic model \eqref{bulkaction}, it would be interesting to check if the CESE conductivity will be corrected by the chiral anomaly. From recent works \cite{Bu:2016oba, Bu:2016vum,Megias:2013joa}, the anomalous corrections to normal transport coefficients start from the second order in the derivative expansion. A study along the line of \cite{Bu:2016oba, Bu:2016vum,Megias:2013joa} will be helpful in clarifying this issue.



\appendix

\section{Technical Details in the Fluid/Gravity Calculations}
\label{appF}

In this appendix, we collect some computational details and useful relations in the fluid/gravity calculations of Section \ref{FluidGravity}.

\subsection{Solving the Bulk Equations} \label{appF-solving}
Following the standard procedure of fluid/gravity correspondence, we first solve constraint equations to derive relations among fluid-dynamical variables. Practically, we find it more convenient to consider certain combinations of constraint and dynamical equations. Below is the listing of solutions to constraint equations,
\begin{align}
 {\Ev}^r=0 ~~&{\Rightarrow}~~ \partial_t Q{\0} + Q_0 (\partial_k u_k{\0})=0, \label{hydro eqs 1st}\\
 {\Ea}^r=0 ~~&{\Rightarrow}~~ \partial_t {{\tQ}}{\0} + {{\tQ}}_0 (\partial_k u_k{\0})=0,\label{hydro eqs 2st}\\
 g^{rr}{\Ew}_{rt}+ g^{rt} {\Ew}_{tt}=0 ~~&{\Rightarrow}~~ 3\partial_t M{\0}+4M_0 (\partial_k u_k{\0})=0,\label{hydro eqs 3st}\\
 g^{rr}{\Ew}_{ri}+g^{rt}{\Ew}_{ti}=0 ~~&{\Rightarrow}~~ \partial_i M{\0}+4 M_0 \partial_t u_i{\0}= 2\sqrt{3} Q_0 F^{\rme}_{it}{\0},\label{hydro eqs 4st}
\end{align}
which are, indeed, the hydrodynamic equations (\ref{conservation1}), expanded to the first-order in the derivative expansion. In obtaining (\ref{hydro eqs 4st}), the first two equations (\ref{hydro eqs 1st})(\ref{hydro eqs 2st}) have been utilized.
To proceed, we turn to dynamical equations and find the corrections in (\ref{correction1}). Dynamical equations will be grouped into the scalar, vector and tensor sectors according to $SO(3)$ symmetry of the boundary spatial directions.

{\bf I. Scalar Sector}. ---
In the scalar sector, we begin with the dynamical equation $E_{rr}=0$:
\begin{equation}
0=r\partial_r^2 h(r)+5\partial_r h(r),
\end{equation}
which is solved by
\begin{equation}
h(r)={C^h_1}+\frac{{C^h_2}}{r^4}.
\end{equation}
The asymptotic condition (\ref{asymp requirement}) requires ${C^h_1}=0$. By Landau-Lifshitz frame condition (\ref{Landau Ttt}), the integration constant ${C^h_2}$ will also be fixed to zero. Therefore,
$h(r)=0$.
From the time components of Maxwell equations ${\Ev}^t=0$ and ${\Ea}^t=0$,
\begin{align}
&r^3\partial_r^2{\av}_t{\br}+3r^2 \partial_r {\av}_t{\br}-4\sqrt{3}Q_0 \partial_r h{\br}=0,\\
&r^3\partial_r^2{\aa_t}{\br}+3r^2 \partial_r {\aa}_t{\br}-4\sqrt{3}{{\tQ}}_0 \partial_r h{\br}=0,
\end{align}
which are solved by
\begin{align}
{\av}_t{\br}={C^{\av}_1}+\frac{{C^{\av}_2}}{r^2}-\frac{2\sqrt{3}Q_0{C^h_2}}{3r^6},\qquad
{\aa}_t{\br}={C^{\aa}_1}+\frac{{C^{\aa}_2}}{r^2}-\frac{2\sqrt{3}{{\tQ}}_0{C^h_2}}{3r^6}.
\end{align}
where ${C^{\av}_1},{C^{\av}_2},{C^{\aa}_1},{C^{\aa}_2}$ are integration constants to be determined by boundary conditions \eqref{asymp requirement}-\eqref{Landau frame}. First, nonzero ${C^{\av}_1},{C^{\aa}_1}$ correspond to non-normalizable modes and would violate the asymptotic requirement (\ref{asymp requirement}). So, ${C^{\av}_1}={C^{\aa}_1}=0$.
The Landau-Lifshitz frame conditions (\ref{Landau Jt}) require ${C^{\av}_2},{C^{\aa}_2}$ to vanish, i.e.
${C^{\av}_2}={C^{\aa}_2}=0$.
The remaining equation of the scalar sector is $g^{rr}E_{rr}+g^{rt}E_{tr}=0$,
\begin{equation}
3\partial_r k{\br}=(2M_0-6r^4) \partial_r h{\br}-24r^3 h{\br}+2\sqrt{3}\big[Q_0\partial_r {\av}_t{\br}+{{\tQ}}_0 \partial_r {\aa}_t{\br}\big]+6r^2 (\partial_k u_k{\0}),
\end{equation}
which is solved by
\begin{equation}
k{\br} =\frac{2}{3}r^3 (\partial_k u_k{\0})+{C^k_1}
+ \frac{2\sqrt{3} (Q_0 {C^{\av}_2}+{{\tQ}}_0 {C^{\aa}_2})} {3r^2} +\left[\frac{2M_0}{3r^4}-\frac{4(Q_0^2+{{\tQ}}_0^2)}{3r^6}\right]{C^h_2}.
\end{equation}
A nonzero ${C^k_1}$ would cause the ${T}_{tt}$ computed from (\ref{Ttt}) to be in contradiction with the Landau-Lifshitz frame convention (\ref{Landau Ttt}). So, ${C^k_1}=0$.
Therefore, all the integration constants in the scalar sector have to be set to zero. The solutions in the scalar sector are summarized as below
\begin{equation}
k{\br}=\frac{2}{3}r^3 (\partial_k u_k{\0}),\qquad
h{\br}=0,\qquad {\av}_t{\br}=0,\qquad {\aa}_t{\br}=0.
\end{equation}

{\bf II. Vector Sector}. ---
Now we consider the helicity one sector, which consists of ${\av}_i{\br},{\aa}_i{\br}, j_i{\br}$ and turns out to be more involved. First consider the Maxwell equations ${\Ev}^i=0$ and ${\Ea}^i=0$
\begin{align} \label{eom vi}
0&=\partial_r\left[r^3f_0(r)\partial_r {\av}_i{\br}\right]+ {2\sqrt{3}Q_0} \partial_r  \Big(\frac{j_i{\br}}{r^4} \Big)
-\frac{\sqrt{3}}{r^2}  \big(\partial_iQ{\0} +Q_0 \partial_t u_i{\0} \big)+{F}_{ti}^{\rme}{\0},\\
 \label{eom ai}
0&=\partial_r\left[r^3f_0(r)\partial_r {\aa}_i{\br}\right]+ {2\sqrt{3}\tQ_0} \partial_r  \Big(\frac{j_i{\br}}{r^4} \Big)
-\frac{\sqrt{3}}{r^2}  \big(\partial_i{{\tQ}}{\0} +{{\tQ}}_0 \partial_t u_i{\0} \big),
\end{align}
which are dynamical equations for ${\av}_i{\br}$ and ${\aa}_i{\br}$, but coupled to $j_i{\br}$. The Einstein equation $W_{ri}=0$ corresponds to dynamical equation for $j_i{\br}$,
\begin{equation} \label{eom ji}
0=r\partial_r^2 j_i{\br}-3\partial_r j_i{\br}+2\sqrt{3}\big[Q_0\partial_r {\av}_i{\br}+{{\tQ}}_0 \partial_r {\aa}_i{\br}\big]+3r^2 \partial_t u_i{\0},
\end{equation}
which couples to ${\av}_i{\br}$ and ${\aa}_i{\br}$.

Our strategy of solving the coupled differential equations \eqref{eom vi}-\eqref{eom ji} is to get rid of $j_i{\br}$ and derive decoupled differential equations for suitably combined variables from ${\av}_i{\br},{\aa}_i{\br}$. To this end, we first integrate over $r$ once in the equation (\ref{eom ji}). As a result,
\begin{equation} \label{eom ji integrated}
0=r\partial_r j_i{\br}-4j_i{\br}+2\sqrt{3}\big[Q_0{\av}_i{\br}+{{\tQ}}_0 {\aa}_i{\br}\big]+r^3 \partial_t u_i{\0},
\end{equation}
where the integration constant is fixed by the Landau-Lifshitz frame convention, i.e., in the near boundary expansion for $j_i{\br}$ the constant should be zero in order to be consistent with \eqref{Landau Ttt}. 

The combinations
$Q_0\times(\ref{eom vi})+{{\tQ}}_0\times(\ref{eom ai})$, 
and ${{\tQ}}_0\times(\ref{eom vi})-Q_0\times(\ref{eom ai})$
give rise to
\begin{align}
0=&\,\partial_r\Big[r^3f_0(r)\partial_r\big(Q_0 {\av}_i{\br}+{{\tQ}}_0 {\aa}_i{\br}\big)\Big]
+ 2\sqrt{3} \big(Q_0^2+{{\tQ}}_0^2\big)\partial_r \Big(\frac{j_i{\br}}{r^4}\Big) \nn \\
&- \frac{\sqrt{3}}{r^2}\Big[Q_0\partial_i Q{\0} +{{\tQ}}_0 \partial_i {{\tQ}}{\0} +\big(Q_0^2 +{{\tQ}}_0^2\big) \partial_t u_i{\0} \Big]
+Q_0{F}_{ti}^{\rme}{\0}, \label{eom vi+ai}
\end{align}
and
\begin{equation}\label{eom vi-ai}
0=\partial_r\left[r^3f_0(r)\partial_r\big({{\tQ}}_0 {\av}_i{\br}-Q_0 {\aa}_i{\br}\big)\right]
- \frac{\sqrt{3}}{r^2}\big({{\tQ}}_0\partial_iQ{\0}-Q_0\partial_i{{\tQ}}{\0}\big) +{{\tQ}}_0 {F}_{ti}^{\rme}{\0}.
\end{equation}
Then, substituting (\ref{eom ji integrated}) into (\ref{eom vi+ai}) yields
\begin{align} \label{vi+ai decoupled}
0=&\,\partial_r\left[r^3f_0(r)\partial_r\big(Q_0{\av_i}{\br}+{{\tQ}}_0{\aa}_i{\br}\big)\right]
-\frac{12}{r^5}\big(Q_0^2+{{\tQ}}_0^2\big)\big(Q_0{\av_i}{\br}+{{\tQ}}_0{\aa}_i{\br}\big) \nn \\
&-\frac{\sqrt{3}}{r^2} \left[3\big(Q_0^2+{{\tQ}}_0^2\big)\partial_tu_i{\0}+ Q_0\partial_i Q{\0}+ {{\tQ}}_0\partial_i {{\tQ}}{\0}\right] + Q_0{F}_{ti}^{\rme}{\0}.
\end{align}

The equation (\ref{eom vi-ai}) can be solved via direct integration over $r$,
\begin{align}
{{\tQ}}_0 {\av}_i{\br}- Q_0 {\aa}_i{\br}
 &= -\int_r^\infty \frac{{\d}{\x} }{{\x} ^3f_0({\x})} \int_{\rh }^{\x} {\d}y
\Big[ \frac{\sqrt{3}}{y^2}\big({{\tQ}}_0 \partial_i Q{\0}- Q_0 \partial_i {{\tQ}}_0\big)-{{\tQ}}_0 {F}_{ti}^{\rme}{\0}\Big] \nn \\
\xrightarrow[]{r\to \infty} &\frac{1}{r}{{\tQ}}_0 {F}_{ti}^{\rme}{\0} -\frac{1}{r^2} \frac{\sqrt{3}}{2\rh }\big({{\tQ}}_0 \partial_i Q{\0}-Q_0\partial_i {{\tQ}}{\0}\big) - \frac{1}{r^2} \frac{1}{2}\rh  {{\tQ}}_0 {F}_{ti}^{\rme}{\0}+\o\(r^{-3}\),
\end{align}
where the lower limit of the inner integral is fixed by regularity (\ref{regularity}) at the unperturbed horizon $r=\rh $ and the upper limit of the outer integral is fixed by asymptotic requirement (\ref{asymp requirement}).

However, the equation (\ref{vi+ai decoupled}) is more complicated and cannot be solved by directly integrating over $r$. Indeed, the homogeneous version of (\ref{vi+ai decoupled}) has two linearly independent solutions given by
$H_1(r){=}r^5f_0^\prime(r)$ and $H_2(r){=}H_1(r) \int_r^\infty  {\d\hr}  \[\hr^3f_0(\hr)H_1\!(\hr)^2\]^{-1}$,
where the second one $H_2(r)$ is obtained by the Liouville formula. Then, one could proceed to solve (\ref{vi+ai decoupled}) by using the method of variation of parameters. In practical calculations, we make a coordinate transformation by $u=\rh /r$ and perform the solving of (\ref{vi+ai decoupled}) in Mathematica. The regularity at the unperturbed horizon and asymptotic requirement completely fix both integration constants. Since the final solution looks quite complicated, we only record the large $r$ behavior for $Q_0{\av_i}{\br}+{{\tQ}}_0{\aa}_i{\br}$,
\begin{align}
Q_0{\av_i}{\br}+{{\tQ}}_0{\aa}_i{\br} \xrightarrow[]{r\to\infty}
& - \frac{1}{r^2} \frac{\sqrt{3} (2\rh ^6+Q_0^2+{{\tQ}}_0^2 )}{4M_0\rh ^3} \left[3 (Q_0^2+{{\tQ}}_0^2 )\partial_tu_i{\0}  + Q_0 \partial_i Q{\0}+ {{\tQ}}_0 \partial_i {{\tQ}}{\0}\right] \nn \\
&+\frac{1}{r}Q_0{F}_{ti}^{\rme}{\0} -\frac{1}{r^2} \Big[\frac{3 (Q_0^2+{{\tQ}}_0^2 )}{4M_0 \rh }+\frac{1}{2}\rh \Big] Q_0 {F}_{ti}^{\rme}+\o\(r^{-3}\).
\end{align}

Therefore, the near boundary behaviors for ${\av}_i{\br}$ and ${\aa}_i{\br}$ are
\begin{align}
{\av}_i{\br}\xrightarrow[]{r\to\infty}&\,- \frac{3\sqrt{3}Q_0} {r^2} \Big(\frac{\rh ^3}{4M_0}+\frac{1}{4\rh }\Big) \partial_tu_i{\0}- \frac{\sqrt{3}}{r^2} \frac{(2\rh ^6+Q_0^2+2{{\tQ}}_0^2)}{4M_0\rh ^3}\partial_iQ{\0}+\frac{\sqrt{3}}{r^2}\frac{Q_0{{\tQ}}_0}{4M_0\rh ^3} \partial_i{{\tQ}}{\0}    \nn \\
& + \frac{1}{r} {F}_{ti}^{\rme}{\0} - \frac{1}{r^2} \Big(\frac{3Q_0^2}{4M_0\rh }+\frac{1}{2}\rh \Big){F}_{ti}^{\rme}{\0}+\o\(r^{-3}\),  \label{large r vi app} \\
 \label{large r ai app}
{\aa}_i{\br}\xrightarrow[]{r\to\infty}&- \frac{3\sqrt{3}{{\tQ}}_0}{r^2} \Big(\frac{\rh ^3} {4M_0}+\frac{1}{4\rh }\Big) \partial_tu_i{\0}- \frac{\sqrt{3}}{r^2} \frac{(2\rh ^6+2Q_0^2+{{\tQ}}_0^2)}{4M_0\rh ^3}\partial_i{{\tQ}}{\0} +\frac{\sqrt{3}}{r^2}  \frac{Q_0{{\tQ}}_0}{4M_0\rh ^3} \partial_iQ{\0}  \nn\\ &
- \frac{1}{r^2} \frac{3Q_0{{\tQ}}_0}{4M_0\rh }{F}_{ti}^{\rme}{\0}+\o\(r^{-3}\).
\end{align}
With large $r$ behaviors of ${\av}_i{\br},{\aa}_i{\br}$ at hand, the equation (\ref{eom ji integrated}) could be solved near the boundary $r=\infty$, yielding
\begin{equation}\label{large r ji app}
j_i{\br}\xrightarrow[]{r\to\infty}r^3\partial_tu_i{\0}+\o (r^{-1}).
\end{equation}

{\bf III. Tensor Sector}.  ---
Finally, the tensor equation ${\Ew}_{ij}-\frac{1}{3}\delta_{ij}{\Ew}_{kk}=0$ gives the dynamical equation for $\alpha_{ij}{\br}$:
\begin{equation}
\partial_r\left[r^5f_0(r)\partial_r \alpha_{ij}{\br}\right]+3r^2\Big[\partial_iu_j{\0}+ \partial_ju_i{\0}-\frac{2}{3}\delta_{ij}(\partial_k u_k{\0}) \Big]=0,
\end{equation}
which can be solved by direct integration over $r$. The final solution for $\alpha_{ij}{\br}$ is
\begin{align} \label{large r alphaij app}
\alpha_{ij}{\br}
 &=  3 \big[\partial_iu_j{\0}+ \partial_ju_i{\0}-\frac{2}{3}\delta_{ij}(\partial_k u_k{\0}) \big] \int_\infty^r \frac{{\d}\x}{\x^5f_0(\x)} \int_{\rh }^{\x} y^2{\d}y,\nn \\
 & \xrightarrow[]{r\to\infty}   \Big(\frac{1}{r}-\frac{\rh ^3}{4r^4}\Big) \Big[\partial_iu_j{\0}+ \partial_ju_i{\0}-\frac{2}{3}\delta_{ij}(\partial_k u_k{\0}) \Big] +\o(r^{-5}).
\end{align}


\subsection{Useful Relations in Deriving Dual Currents} \label{appF-relations}

In this appendix, we summarize some formulas that are quite lengthy but useful towards deriving non-anomalous parts of the currents in \eqref{covariantJm}\eqref{covariantJm5}.
From \eqref{Jti}\eqref{J5ti}, armed with the near-boundary behaviors derived in appendix \ref{appF-solving}, the vector and axial currents of the boundary theory are
\begin{align} \label{Jt+J5t pre-covariant}
{J}^t &= 2\sqrt{3}Q ,\qquad \qquad {J}^t_5=2\sqrt{3}{{\tQ}} ,\\
{J}^i &=\,\sqrt{3}Q_0u_i -3\sqrt{3}Q_0\Big(\frac{\rh ^3}{2M_0}+\frac{1}{2\rh }\Big) \partial_tu_i{\0}-\frac{\sqrt{3} (2\rh ^6+Q_0^2+2{{\tQ}}_0^2 )}{2M_0\rh ^3} \partial_iQ{\0}  \nn \\ &
~~~+\frac{\sqrt{3}Q_0{{\tQ}}_0}{2M_0\rh ^3}\partial_i{{\tQ}}{\0}- \Big(\frac{3Q_0^2} {2M_0\rh }+\rh \Big){F}_{ti}^{\rme}{\0},  \label{Ji pre-covariant} \\
J_5^i &=\,\sqrt{3}{{\tQ}}_0u_i -3\sqrt{3}{{\tQ}}_0\Big(\frac{\rh ^3}{2M_0}+ \frac{1}{2\rh }\Big) \partial_tu_i{\0}-\frac{\sqrt{3} (2\rh ^6+2Q_0^2+{{\tQ}}_0^2  )}{2M_0\rh ^3} \partial_i{{\tQ}}{\0}   \nn \\ &
~~~+\frac{\sqrt{3}Q_0{{\tQ}}_0}{2M_0\rh ^3}\partial_iQ{\0}- \frac{3Q_0{{\tQ}}_0} {2M_0\rh } {F}_{ti}^{\rme}{\0}, \label{J5i pre-covariant}
\end{align}
where $\partial_tu_i{\0}$ will be replaced via the constraint relation (\ref{hydro eqs 4st})
\begin{equation}
\partial_tu_i{\0}=-\frac{\partial_iM{\0}}{4M_0}-\frac{\sqrt{3}Q_0}{2M_0}{F}_{ti}^{\rme}{\0}.
\end{equation}
Meanwhile, the derivative $\partial_i M$ would be replaced by
\begin{equation}
\partial_iM{\0}=\frac{1}{\rh ^3}\big(Q_0\partial_iQ{\0}+{{\tQ}}_0 \partial_i {{\tQ}}{\0}\big)
+2\rh ^3 \bigg(1-\frac{Q_0^2+{{\tQ}}_0^2}{2\rh ^6}\bigg) \partial_i\rh {\0},
\end{equation}
which is obtained by expanding
\begin{equation}
1-\frac{M }{\rh ^4 }+\frac{Q^2 +{{\tQ}}^2 }{\rh^6 }=0,
\end{equation}
around $x^\mu=0$ up to first-order in derivative expansion.

Then, the derivative terms in \eqref{Ji pre-covariant}\eqref{J5i pre-covariant} are linear combinations of $\partial_i \rh$, $\partial_i Q$ and $\partial_i \tQ$, which have to be re-parameterized in terms of derivatives of fluid-dynamical variables. Via the relations \eqref{define chemical potential}\eqref{temperature field}, it is straightforward although tedious to derive the following expressions
\begin{align} \label{temperature derivative}
\frac{\partial_i T}{T}=-\frac{2\big(Q_0\partial_iQ{\0}+{{\tQ}}_0\partial_i {{\tQ}}{\0}\big)}{2\rh ^6-Q_0^2-{{\tQ}}_0^2}
+\frac{2\rh^6+5\big(Q_0^2+{{\tQ}}_0^2 \big)}{\big(2\rh^6-Q_0^2-{{\tQ}}_0^2\big)} \frac{\partial_i\rh {\0}}{\rh },
\end{align}
\begin{align} \label{mu derivative}
\partial_i\Big(\frac{\mu}{T}\Big)&=
\frac{2\sqrt{3}\pi \rh ^2  \big(\rh \partial_i{{Q}}{\0}- 3{{Q}}_0 \partial_i \rh {\0}\big)}{2\rh ^6-Q_0^2- {{\tQ}}_0^2}
+ \frac{4\sqrt{3}\pi {{Q}}_0\rh ^2  \big(\rh Q_0\partial_iQ{\0}+\rh {{\tQ}}_0 \partial_i{{\tQ}}{\0} -3Q_0^2 \partial_i \rh {\0}-3{{\tQ}}_0^2\partial_i\rh {\0}\big)}{\big(2\rh ^6-Q_0^2-{{\tQ}}_0^2\big)^2},\\
\label{mu5 derivative}
\partial_i\Big(\frac{\mu_5}{T}\Big)&=
\frac{2\sqrt{3}\pi \rh ^2  \big(\rh \partial_i{{\tQ}}{\0}- 3{{\tQ}}_0 \partial_i \rh {\0}\big)}{2\rh ^6-Q_0^2- {{\tQ}}_0^2}
+ \frac{4\sqrt{3}\pi {{\tQ}}_0\rh ^2  \big(\rh Q_0\partial_iQ{\0}+\rh {{\tQ}}_0 \partial_i{{\tQ}}{\0} -3Q_0^2 \partial_i \rh {\0}-3{{\tQ}}_0^2\partial_i\rh {\0}\big)}{\big(2\rh ^6-Q_0^2-{{\tQ}}_0^2\big)^2},
\end{align}
which could be inverted to express $\partial_i \rh$, $\partial_i Q$ and $\partial_i \tQ$ in terms of $\partial_i T$, $\partial_i(\mu/T)$ and $\partial_i(\mu_5/T)$.

\section{Technical Details in Linear Response Analysis}
\label{appL}

In this appendix, we collect calculation details in the linear response analysis presented in Section \ref{LinearAnalysis}.

{\bf I. Identify External Sources}. --- According to the holographic dictionary, non-normalizable modes of bulk fields act as external sources. Thus, near the conformal boundary $r\to\infty$, we require
\begin{align} \label{asymptotic metric}
\delta g_{ti}(r,t) &=r^2 \big[{\dhti}(t)+ \o(r^{-1}) \big],\\
{\delta{\Av}_i}(r,t) &= {\davi}(t)+ \o(r^{-1}),\\
{\delta{\Aa}_i}(r,t) &= {\daai}(t)+ \o (r^{-1}). \label{asymptotic gauge fields}
\end{align}
As a result, the boundary metric is perturbed to be $\eta_{\mu\nu}+{\dhti}(t)$. The purpose of introducing a boundary metric perturbation ${\dhti}(t)$ is to turn on a thermal gradient $\nabla_i T$. Indeed, via the diffeomorphism invariance, one could show that a thermal gradient $\nabla_i T$ leads to a boundary metric perturbation ${\dhti}(t)$ \cite{Hartnoll:2009sz,Herzog:2009xv},
\begin{equation} \label{thermal-metric}
i\omega {\delta h_{tj}^{(0)}}=-\frac{\nabla_j T}{T}.
\end{equation}
On the other hand, a thermal gradient $\nabla_i T$ also induces perturbations to boundary vector and axial gauge potentials,
\begin{equation} \label{thermal-gauge}
i\omega \delta {\av}_j^{(\textrm{T})} =\mu \frac{\nabla_j T}{T},\qquad
i\omega \delta {\aa}_j^{(\textrm{T})} =\mu_5 \frac{\nabla_j T}{T},
\end{equation}
where the superscript $^{(\textrm{T})}$ is to emphasize that the potential perturbations in (\ref{thermal-gauge}) are generated by a thermal gradient and will be vanishing once the thermal gradient is turned off. The chemical potentials $\mu,\mu_5$ are defined as in \eqref{define chemical potential}.
Consequently, we identify external vector and axial electric fields as
\begin{equation} \label{vector/axial E}
\begin{split}
E_i&=-\partial_t\big({\davi}- \delta{\av}_i^{(\textrm{T})}\big)=i\omega \big({\davi}+\mu {\dhti}\big),\\
E_i^5&=-\partial_t\big({\daai}- \delta{\aa}_i^{(\textrm{T})}\big)=i\omega \big({\daai}+\mu_5 {\dhti}\big).
\end{split}
\end{equation}
In \eqref{thermal-metric}-\eqref{vector/axial E} we have assumed plane wave ansatz for external perturbations
\begin{equation}
{\dhti}(t)\sim e^{-i\omega t}{\dhti}(\omega),\qquad
{\davi}(t)\sim e^{-i\omega t}{\davi}(\omega),\qquad
{\daai}(t)\sim e^{-i\omega t}{\daai}(\omega).
\end{equation}

{\bf II. Bulk Equations Linearized}. --- To linear order in perturbations, the bulk equations of motion \eqref{Einstein}-\eqref{MaxwellA} become
\begin{align}
{\Ew}_{ri} =0~~{\Rightarrow}~~0&= r^3 \partial_r^2 \delta g_{ti} +r^2 \partial_r \delta g_{ti} -4r \delta g_{ti}+ 2\sqrt{3} \partial_r\big(Q {\delta{\Av}_i}+ {{\tQ}} {\delta{\Aa}_i}\big), \label{Eri}\\
{\Ew}_{ti} =0~~{\Rightarrow}~~0&= \,r^5f(r) \partial_r^2 \delta g_{ti}+r^4f(r) \partial_r \delta g_{ti} +r^3 \partial_r \partial_t \delta g_{ti}-2r^2 \partial_t \delta g_{ti}\nn \\
&\quad - 4 r^3 f(r) \delta g_{ti}+2\sqrt{3}\left[r^2 f(r) \partial_r+\partial_t\right] \big(Q {\delta{\Av}_i} +{{\tQ}} {\delta{\Aa}_i}\big), \label{Eti}
\end{align}
as well as
\begin{align}
{\Ev}^i=0 ~~{\Rightarrow}~~ 0=&\,r^3f(r)\partial_r^2 {\delta{\Av}_i} +\partial_r \left[ r^3f(r)\right] \partial_r {\delta{\Av}_i} + 2r\partial_r \partial_t {\delta{\Av}_i} + \partial_t {\delta{\Av}_i}  \nn \\ &
+2\sqrt{3} Q r^{-3} \left(r \partial_r \delta g_{ti}- 2 \delta g_{ti}\right), \label{EVi}\\
{\Ea}^i=0 ~~{\Rightarrow}~~ 0=&\,r^3f(r)\partial_r^2 {\delta{\Aa}_i} +\partial_r \left[ r^3f(r)\right] \partial_r {\delta{\Aa}_i} + 2r\partial_r \partial_t {\delta{\Aa}_i} + \partial_t {\delta{\Aa}_i}  \nn \\ &
+ 2\sqrt{3} {{\tQ}} r^{-3} \left(r \partial_r \delta g_{ti}- 2 \delta g_{ti}\right).\label{EAi}
\end{align}
The combination $r^2f(r){\Ew}_{ri}-{\Ew}_{ti}=0$ results in a simpler equation
\begin{align} \label{Eri-Eti}
0=r^3\partial_r\delta g_{ti}-2r^2\delta g_{ti}+2\sqrt{3}  ({{\tQ}}{\delta{\Aa}_i}+Q{\delta{\Av}_i} ),
\end{align}
which helps to decouple ${\delta{\Av}_i}$, ${\delta{\Aa}_i}$ from $\delta g_{ti}$. As a result,
\begin{align}
0&=\partial_r\left[r^3f(r)\partial_r {\delta{\Av}_i}\right]+ 2r \partial_r\partial_t {\delta{\Av}_i}+ \partial_t {\delta{\Av}_i}-12Q r^{-5}  (Q {\delta{\Av}_i}+ {{\tQ}} {\delta{\Aa}_i}  ), \label{del VA1} \\
0&=\partial_r\left[r^3f(r)\partial_r {\delta{\Aa}_i}\right]+ 2r \partial_r\partial_t {\delta{\Aa}_i} + \partial_t {\delta{\Aa}_i}-12 {{\tQ}} r^{-5}  (Q {\delta{\Av}_i}+ {{\tQ}} {\delta{\Aa}_i}  ).  \label{del VA2}
\end{align}

To proceed, we define the variables
\begin{equation}
\begin{split}
X_i&\equiv Q{\delta{\Av}_i}+ {{\tQ}} {\delta{\Aa}_i}- ( Q{\davi}+ {{\tQ}} {\daai} ),\\
Y_i&\equiv {{\tQ}}{\delta{\Av}_i}- Q {\delta{\Aa}_i} - ( {{\tQ}}{\davi}- Q {\daai} ),
\end{split}
\end{equation}
so that $\lim_{r\to\infty}X_i\to0$, and $\lim_{r\to\infty} Y_i\to 0$.
Then, the equations \eqref{del VA1}\eqref{del VA2} turn into decoupled equations for $X_i$ and $Y_i$,
\begin{align} \label{eom Xi}
0 &= \partial_r\left[r^3f(r)\partial_r X_i\right]+2r\partial_r\partial_t X_i+\partial_t X_i-12  \big(Q^2+{{\tQ}}^2 \big)r^{-5}X_i \nn \\
&\quad +\big[\partial_t-12 (Q^2+{{\tQ}}^2 )r^{-5}\big]  \big(Q{\davi}+ {{\tQ}} {\daai} \big),\\
\label{eom Yi}
0 &=\partial_r\left[r^3f(r)\partial_r Y_i\right]+2r\partial_r\partial_t Y_i+\partial_t Y_i+\partial_t \big({{\tQ}}{\davi}-Q {\daai}\big).
\end{align}
Inspired by the structure of source terms in \eqref{eom Xi}\eqref{eom Yi}, we could factorize $X_i$ and $Y_i$ as
\begin{equation} \label{basis decomposition}
\begin{split}
X_i(r,t)&=S_1(r,\partial_t)\big[Q{\davi}(t)+ {{\tQ}} {\daai}(t) \big],\\
Y_i(r,t)&=S_2(r,\partial_t)\big[Q{\davi}(t)+ {{\tQ}} {\daai}(t) \big].
\end{split}
\end{equation}

In Fourier space by $\partial_t\to -i\omega$, we have
$
S_1(r,\partial_t)\to S_1(r,\omega),
S_2(r,\partial_t)\to S_2(r,\omega).
$
Eventually, dynamical equations for ${\delta{\Av}_i}, {\delta{\Aa}_i}$ turn into decoupled ordinary differential equations for scalar functions $S_1,S_2$:
\begin{align}\label{eom S1}
0 &=\partial_r\left[r^3f(r)\partial_r S_1\right] -2i\omega r\partial_r S_1- i\omega S_1 -12  (Q^2+{{\tQ}}^2 ) r^{-5} S_1 -i\omega -12  (Q^2+{{\tQ}}^2 ) r^{-5},
\\  \label{eom S2}
0 &=\partial_r\left[r^3f(r)\partial_rS_2\right] -2i\omega r\partial_r S_2-i\omega S_2 -i\omega.
\end{align}
The asymptotic expansions of ${\delta{\Av}_i}$ and ${\delta{\Aa}_i}$ in \eqref{asymptotic delV}\eqref{asymptotic delA} get translated into near boundary behavior for $S_1,S_2$, which we sketchily summarise as
\begin{equation} \label{asymp expansion S1S2}
S_i\xrightarrow[]{r\to\infty} \frac{s_i^{(1)}}{r}+\frac{s_i}{r^2}+\frac{\log r}{r^2} s_i^{\textrm{L}},\qquad\qquad i=1,2,
\end{equation}
where $s_i^{(1)},s_i^{\rm L}$ are easily read off from \eqref{asymptotic delV}\eqref{asymptotic delA} and $s_i$ will be solved in the following. Afterwards, we determine the frequency-dependence of conductivities $\sigma$ and $\saq$.

We turn to a bounded radial coordinate by the transformation,
\begin{equation}
u\equiv \frac{\rh }{r} \Rightarrow u\in[0,1],
\end{equation}
so that the conformal boundary is located at $u=0$ and the event horizon  is at $u=1$. In the $u$-coordinate, equations \eqref{eom S1}\eqref{eom S2} are turned into
\begin{equation} \label{eom S1S2}
\begin{split}
&0=u^2\partial_u\left[u^{-1}f(u)\partial_uS_1\right]+2i{\homega} u \partial_uS_1- i{\homega}(S_1+1)-4\left({\hmu}^2+{\hmu}_5^2\right)u^5\left(S_1+1\right),\\
&0=u^2\partial_u\left[u^{-1}f(u)\partial_uS_2\right]+2i{\homega}u\partial_uS_2- i{\homega}\left(S_2+1\right),
\end{split}
\end{equation}
where
\begin{equation}
{\homega}=\frac{\omega}{\rh },~~~~~~~{\hmu}=\frac{\mu}{\rh },~~~~~~{\hmu}_5= \frac{\mu_5}{\rh },~~~~~~~f(u)=1-u^4+\frac{1}{3}\left({\hmu}^2+{\hmu}_5^2\right) u^4(u^2-1).
\end{equation}
The equations (\ref{eom S1S2}) will be first solved analytically in the hydrodynamic limit to compare with the results of Section \ref{FluidGravity} and then numerically in order to reveal the frequency dependence of conductivities.

{\bf III. Hydrodynamic Limit}. ---
First, we consider the hydrodynamic limit where $ \homega \ll 1$ so that we could compute $\sigma$ and $\saq$ analytically. Introduce a formal parameter by the rescaling
$
{\homega}\to \lambda {\homega}$.
Then, $S_1,S_2$ are expanded as
$S_1=\sum_{n=0}^\infty \lambda^n S_1^{[n]}, S_2=\sum_{n=0}^\infty \lambda^n S_2^{[n]}$.
To the zeroth order $\o(\lambda^0)$,
\begin{equation}
S_1^{[0]}=-\frac{3\left({\hmu}^2+{\hmu}_5^2\right)}{2\left(3+{\hmu}^2+ {\hmu}_5^2\right)}u^2,\qquad\qquad S_2^{[0]}=0.
\end{equation}
To the first-order $\o(\lambda^1)$, $S_2^{[1]}$ can be solved by direct integration over $r$ and the result is
\begin{equation}
S_2^{[1]}=i{\homega}\int_0^u \frac{\x\d\x}{f(\x)}\int_1^{\x} \frac{\d y}{y^2} \xrightarrow[]{u\to 0} \frac{1}{2}i {\homega} \left(-2u+u^2\right) +\o (u^3).
\end{equation}
The equation for $S_1^{[1]}$ is
\begin{equation}
0=u^2\partial_u\left[u^{-1}f(u)\partial_u S_1^{[1]}\right]-4\left({\hmu}^2+ {\hmu}_5^2 \right)u^5 S_1^{[1]}+2 i{\homega} u \partial_u S_1^{[0]} -i{\homega} S_1^{[0]} - i{\homega},
\end{equation}
which we solved by using Mathematica's DSolve command. The integration constants are fixed by regularity at $u=1$ and asymptotic requirement at the boundary $u=0$. Given that the final solution for $S_1^{[1]}$ is quite lengthy, we record its near boundary expansion only, 
\begin{equation}
S_1^{[1]}\xrightarrow[]{u\to 0}-i{\homega} u +i {\homega} u^2 \frac{\left(-6+{\hmu}^2+{\hmu}_5^2\right)}{\left(3+{\hmu}^2+ {\hmu}_5^2\right)}+ \o(u^3),
\end{equation}
which is enough for calculating the conductivities.

Recall the AC Conductivities \eqref{AC Conductivities} and the near boundary behaviors in \eqref{asymp expansion S1S2}, the small $\omega$ limits of conductivities $\sv\bo$,$\sa\bo$ are
\begin{align} \label{sigma DC1}
\sv\bo&=\frac{i}{\omega} \frac{3Q^2 \rh ^2}{\rh ^6+Q^2+{{\tQ}}^2}+\frac{ Q^4+ Q^2  (5{{\tQ}}^2-4\rh ^6 )+ 4 ({{\tQ}}^2+ \rh ^6 )^2} {4 (\rh ^6+Q^2+{{\tQ}}^2 )^2}\rh  +\cdots,\\
\label{sigma5 DC1}
\sa\bo&=\frac{i}{\omega} \frac{3Q {{\tQ}} \rh ^2}{\rh ^6+Q^2+{{\tQ}}^2} -\frac{3Q{{\tQ}}  (4\rh ^6+Q^2+{{\tQ}}^2 )}{4 (\rh ^6+Q^2+{{\tQ}}^2  )} \rh  + \cdots,
\end{align}
where $\cdots$ denote higher powers in $\omega$ corrections. Results above are in perfect agreement with the linear response limit of fluid/gravity calculations \eqref{sigmaVo}\eqref{sigmaAo} by utilizing the following relation $ M=\rh ^4+({Q^2+{{\tQ}}^2})/{\rh ^2}$,
as well as taking into account \eqref{sigma=sigmaQ+sigma0}.

{\bf IV. Numerical Technique}. ---
For generic frequency $\omega$, we were able to solve ODEs in (\ref{eom S1S2}) numerically only. Inspired by the expressions \eqref{AC Conductivities} 
we turn to variables
\begin{equation}
S_+=\frac{{\hmu}^2 S_1+ {\hmu}_5^2 S_2}{{\hmu}^2+{\hmu}_5^2}, \qquad\qquad
S_-= \frac{{\hmu} {\hmu}_5(S_1-S_2)}{{\hmu}^2+{\hmu}_5^2},
\end{equation}
which obey coupled ODEs,
\begin{equation} \label{S+S-}
\begin{split}
&0=u^2\partial_u\left[u^{-1}f(u)\partial_u S_+\right] +2i{\homega} u\partial_u S_+ -i {\homega} (S_+ + 1)-4u^5 {\hmu} \left({\hmu} S_+ + {\hmu}_5 S_- +{\hmu}\right),\\
&0=u^2 \partial_u \left[u^{-1} f(u) \partial_u S_- \right] +2i{\homega} u\partial_u S_- - i{\homega} S_- - 4u^5 {\hmu}_5 \left({\hmu} S_+ + {\hmu}_5 S_- + {\hmu}\right).
\end{split}
\end{equation}

We numerically solve (\ref{S+S-}) within the pseudospectral collocation method.
The boundary conditions for $S_+,S_-$ could be straightforwardly obtained from those for $S_1,S_2$. For the sake of performing numerical calculations within the spectral collocation method, we summarize the boundary conditions as equalities
\begin{equation}
S_+=S_-=0,\qquad  {\rm at} \qquad u=0,
\end{equation}
\begin{align}
\Big\{\big[-4+\frac{2}{3} ({\hmu}^2+{\hmu}_5^2 )+2i{\homega}\big] \partial_uS_+ -i{\homega} (S_+ + 1)- 4 {\hmu} \left({\hmu} S_+ + {\hmu}_5 S_- + {\hmu} \right)\Big\}\Big|_{u=1}&=0,\\
\Big\{\big[-4+\frac{2}{3} ({\hmu}^2+{\hmu}_5^2 )+2i{\homega}\big] \partial_u S_- -i{\homega} S_- - 4 {\hmu}_5 \left({\hmu} S_+ + {\hmu}_5 S_- + {\hmu}\right)\Big\}\Big|_{u=1}&=0.
\end{align}
With equations (\ref{S+S-}) solved, AC conductivities $\sigma,\sigma_5$ are extracted from the near boundary behavior of $S_+$ and $S_-$:
\begin{equation}
\sv\bo=\frac{\rh ^2}{i\omega} \partial_u^2 \big(S_+ + \frac{1}{2} \omega^2 u^2\log{u} \big)\Big|_{u=0}-\frac{1}{2}i\omega,
\qquad \qquad
\sa\bo=\frac{\rh ^2}{i\omega} \partial_u^2 S_- \Big|_{u=0}.
\end{equation}

\section*{Acknowledgements}
{
We would like to thank Ya-Peng Hu, Kyung Kiu Kim, Keun-Young Kim, Li Li, Wei-Jia Li, Yan Liu, Shi Pu,  Ya-Wen Sun, and Di-Lun Yang for useful discussions, as well as the anonymous referee for helpful suggestions.
Y. Bu is supported by the Fundamental Research Funds for the Central Universities under grant No.122050205032 and the Natural Science Foundation of China (NSFC) under the grant No.11705037.
R. G. Cai is supported by the NSFC (No.11690022, No.11375247, No.11435006, No.11647601), Strategic Priority Research Program of CAS (No.XDB23030100), Key Research Program of Frontier Sciences of CAS.
Q. Yang is supported by the Beijing Normal University Grant (No.312232102) and China Postdoctoral Science Foundation Funded Project (No.212400210).
Y. L. Zhang is supported by the Young Scientist Training Program in APCTP, which is funded by the Ministry of Science, ICT and Future Planning(MSIP), Gyeongsangbuk-do and Pohang City.
}


\end{document}